\documentstyle[preprint,floats,prl,aps]{revtex}
\input epsf.tex
\def\eqn{\begin{equation}}
\def\endeqn{\end{equation}}
\def\eqna{\begin{eqnarray}}
\def\endeqna{\end{eqnarray}}
\begin{document}
\draft
\preprint{CLNS95/1382,~~ hep-ph/9512327}
\title
{
Radiative Corrections to the Muonium Hyperfine Structure.\\  
I.  The $\alpha^2 (Z\alpha )$ Correction}
\author{T. Kinoshita\thanks{e-mail: tk@hepth.cornell.edu}
and M. Nio\thanks{e-mail: makiko@pa.uky.edu}
\thanks{ present address: Department
of Physics and Astronomy,
University of Kentucky, Lexington, KY 40506
} }

\address{ Newman Laboratory of Nuclear Studies,
Cornell University, Ithaca, NY 14853 }
\date{\today}
\maketitle
\begin{abstract}
This is the first of a series of papers on 
a systematic application of the NRQED bound state theory of Caswell and 
Lepage to higher-order radiative corrections
to the hyperfine structure of the muonium ground state.
This paper describes  the calculation of
the $\alpha^2 (Z\alpha)$ radiative correction.
Our result for the complete $\alpha^2 (Z\alpha)$ correction is
0.424(4) kHz,
which reduces the theoretical uncertainty significantly.
The remaining uncertainty is dominated by that of the numerical
evaluation of the nonlogarithmic part of the $\alpha (Z\alpha )^2$
term and logarithmic terms of order $\alpha^4$.
These terms will be treated in the subsequent papers.
\end{abstract}

\vspace{5ex}

%\narrowtext
\section{Introduction }
\label{sec:INTRO}

The hyperfine structure of hydrogenic atoms is one of the well-understood
problems both experimentally and theoretically.
Especially, the muonium has played an important role in the
precision test of QED because its radiative corrections have been
calculated to high orders and  its
hyperfine splitting has been measured very precisely~\cite{mariam}:
\eqn
  \Delta \nu (\mbox{exp})~ =~ 4~463~302.88~(16)~\mbox{kHz}~~~~~~
(0.036~\mbox{ppm}).        \label{meas}
\endeqn
Furthermore, a new experiment is in progress to improve the precision
of $\Delta \nu (\mbox{exp})$ to about 0.007 ppm
\cite{hughes}.
To match this experimental accuracy, it is necessary to improve the
theory of the $\alpha^2 (Z\alpha)$ and $\alpha (Z\alpha )^2$ non-recoil
radiative corrections as well as the leading $\ln(Z\alpha )$ terms of
order $\alpha^{4-n} (Z\alpha )^n$, n = 1, 2, 3, and some relativistic
corrections.
This paper presents details of the calculation of  the $\alpha^2(Z\alpha)$
radiative correction.
A preliminary report of this work has been published \cite{KN1}.

As is well known, the bulk of the hyperfine splitting can be explained
by the nonrelativistic quantum mechanics and is given
by the Fermi formula  \cite{fermi}
\eqn
  E_F={16 \over 3}\alpha^2 c R_{\infty} {{m} \over {M}}
  \left [ 1+{m \over M} \right ]^{-3},     \label{EF}
\endeqn
where 
$R_{\infty}$ is the Rydberg constant for infinite nuclear mass,
and $m$ and $M$ are the electron and muon masses, respectively.

Many correction terms have been calculated over several decades
since the pioneering work of Fermi. Unfortunately, different terms were
often evaluated by different methods making comparison of the results
nontrivial in some cases.
This causes a particularly difficult problem in identifying and
evaluating higher order
correction terms. Recently, however, Lepage and his collaborators have
developed an
approach, called NRQED, to deal with the nonrelativistic and weakly coupled
bound systems
consistently, starting from quantum electrodynamics
(QED) \cite{CL,KL,positronium}.
This provides a solid framework for evaluating higher order
radiative corrections  systematically and unambiguously.  
This series of papers deal with a  treatment of radiative corrections 
of the muonium hyperfine structure within the framework of NRQED.

Before describing our calculation, let us summarize the previous results
on the muonium hyperfine splitting $\Delta \nu$.
It is customary to classify the QED corrections to $\Delta \nu$  
into three types:
radiative non-recoil correction, pure recoil correction,
and radiative-recoil correction.
We use the convention such that electron charge is $e$ and 
the charge of the positive muon is $-Ze$.
Of course $Z = 1$ for the muon, but it is kept
in the formula in order to identify the origin of corrections.
Note that each radiative photon on the electron-line contributes a factor
$\alpha$,
that on the muon line a factor $Z^2 \alpha$,
and one jumping from electron to muon a factor $Z\alpha$.
This factor also arises from the effect of binding on the velocity distribution 
of atomic electrons.
In addition, there are small corrections due to 
the hadronic vacuum polarization 
and    weak interaction effects.
Thus one may write
\eqna
  \Delta \nu (\mbox{theory}) &=&\Delta \nu (\mbox{rad})
  + \Delta \nu (\mbox{recoil})
  + \Delta \nu (\mbox{rad-recoil})
\nonumber \\
  &+& \Delta \nu (\mbox{hadron})
  + \Delta \nu (\mbox{weak}) .    \label{theoryform}
\endeqna

Purely radiative terms of orders $\alpha (Z\alpha )$ and
$\alpha (Z \alpha)^2$ have been known for some time~\cite{SY}:
\eqna
  \Delta \nu (\mbox{rad})&=& (1 + a_{\mu} ) \left (  1 + {3 \over 2}
(Z \alpha )^2 + a_e + \alpha (Z \alpha )( \ln 2 - {5 \over 2} ) \right .
 \nonumber   \\
  & &\left . - {{8 \alpha (Z \alpha )^2} \over {3 \pi}} \ln (Z\alpha)
\left [ \ln (Z\alpha) - \ln 4 + {281 \over 480} \right ] \right .
 \nonumber  \\
  & &\left . + {{\alpha (Z\alpha )^2} \over {\pi}} (14.88 \pm 0.29 )
\right ) E_F.  \label{nonrecoil}
\endeqna
Here $a_e$ and $a_{\mu}$ are the anomalous magnetic moments of the
electron and muon, respectively.
The appearance of the factor $(1 + a_{\mu} )$ in (\ref{nonrecoil}) is
in accord with our definition of $E_F$ in (\ref{EF}).
Note that the number 14.88  in the $\alpha(Z\alpha)^2$ correction is
different from 15.39 reported in Ref. \cite{SY}. This is due to
the recent discovery of two mistakes in the literature.  
The first error is in the  calculation of the $\alpha(Z\alpha)^2 $ correction
due to
the vacuum polarization insertion in the transverse  photon in Ref. \cite{BE}.
Recently several people independently found  \cite{thesis,sapirstein2,SGS}
that this contribution
is $E_F\alpha(Z\alpha)^2/\pi ( -4/5)$,
not $E_F\alpha(Z\alpha)^2/\pi ( -2/3)$ given in Ref.\cite{BE}.
The second error was caused by  omission of a part of the contribution due to
the vacuum polarization insertion in the Coulomb  photon,
$E_F\alpha(Z\alpha)^2/\pi ( -8/15) \ln 2$,
when it was combined with the
radiative photon contribution  $E_F\alpha(Z\alpha)^2/\pi (15.10\pm 0.29)$
\cite{sapirstein2}.

The known recoil corrections add up to~\cite{SY}
\eqna
  \Delta \nu (\mbox{recoil})&=& \left ( - {{3Z\alpha} \over {\pi}}
{{m M} \over {M^2 -m^2 }} \ln {M
\over {m}} \right . \nonumber \\
  & &\left .  + {{\gamma^2} \over {m M}} \left
[ 2 \ln {{m_r} \over {2\gamma}} - 6 \ln 2 + {65 \over 18}
\right ] \right ) E_F ,        \label{recoil}
\endeqna
where $\gamma \equiv Z \alpha m_r$, $m_r =  m M
/(m + M )$.
The radiative-recoil contributions, which arise from both electron and muon
lines
and from vacuum polarizations, are given by 
\footnote{Eq. (6) of Ref. \cite{KN1} 
is valid only for $Z = 1$ although it does not affect the muonium.  
We thank B. N. Taylor and P. Mohr for 
pointing out this oversight.}
\eqna
  \Delta \nu (\mbox{rad-recoil})&=& {{\alpha (Z\alpha)} \over
\pi^2} {m \over M} \left ( - 2 \ln^2 {M \over {m}}
+ {13 \over 12} \ln {M \over m}  + 6 \zeta (3) + \zeta (2) - {71 \over 72} + 3
\pi^2 \ln 2 \right . \nonumber \\
 & &\left .  + Z^2 \left [ {9 \over 2} \zeta (3) + {39 \over 8} - 3\pi^2 \ln 2
\right ]   \right .  \nonumber   \\
  & &\left . + {\alpha \over \pi} \left [ - {4 \over 3} \ln^3 {{M}
\over m} + {4 \over 3} \ln^2 {{M} \over m}  + {\cal O}
\left ( \ln {{M} \over m} \right ) \right ] \right ) E_F ~.
    \label{radrecoil}
\endeqna
The $\alpha (Z\alpha )(m/M)$  and $Z^2\alpha (Z\alpha )(m/M)$ terms are known
exactly~\cite{SY,EKS0}.
The $\ln^3$ and $\ln^2$ parts of the $\alpha^2 (Z\alpha )$ term were
evaluated by Eides $et~al$. \cite{EKS1}.

The hadronic  vacuum polarization 
contributes  \cite{KF}
\eqna
  \Delta \nu (\mbox{hadron})&=& {{\alpha (Z\alpha)} \over
\pi^2} {m M \over m_{\pi}^2}  ( 3.75 \pm 0.24) E_F 
\nonumber \\
&=& 0.250~(16) ~\mbox{kHz}~, ~~~~~~~\label{hadron}  
\endeqna
where $m_{\pi}$ is the charged pion mass.

%%%%%%
\begin{table}
\caption{ Contributions of various terms to the hyperfine splitting 
of the ground state muonium. 
(The new result of this paper is not included.) 
They are represented in units of  kHz.  
The contribution from the muon anomalous magnetic moment is included
in each non-recoil radiative correction term in the left column.
% \multicolumn{2}{c}{ radiative non-recoil} 
%\multicolumn{2}{c}{ recoil and radiative recoil}
\label{table0}}
\[
\begin{array}{crccr}
\\ \hline 
{\rm term}             &   {\rm kHz}  &~~~~~&   {\rm term}   &   {\rm  kHz}
\\ \hline 
E_F             &    4~459~032.409  &~~~~~&   Z\alpha m/M    & -800.304
\\ \hline
a_e             &        5~170.927  &~~~~~&   (Z\alpha)^2m/M &    8.982
\\ \hline
(Z\alpha)^2     &          356.174  &~~~~~&   \alpha(Z\alpha) m/M
                                                            &   -2.636  
\\ \hline
\alpha(Z\alpha) &         -429.036  &~~~~~&  Z^2\alpha(Z\alpha)m/M
                                                            &   -1.190
\\ \hline
\alpha(Z\alpha)^2\ln^2(Z\alpha)^{-1}
                &          -35.606  &~~~~~&   \alpha^2(Z\alpha)m/M 
                                                            &   -0.044
\\ \hline
\alpha(Z\alpha)^2 \ln(Z\alpha)^{-1}
                &           -5.796  &~~~~~&  {\rm  hadron}   &    0.250
\\ \hline
\alpha(Z\alpha)^2
                &            8.207 &~~~~~&  {\rm  weak  }    &   -0.065
\\ \hline
\end{array}
\]
\end{table}
%%%%%%%

Finally there is a small contribution due to the $Z^0$ exchange.
Our re-evaluation of the standard-model estimate~\cite{BF,BFer} gives
\footnote{ This is in agreement with the corrected value given in 
Ref.\cite{BFer} and has a sign opposite to that of Ref. \cite{KN1}. 
The same result was also obtained by J. R. Sapirstein and 
by M. I. Eides. 
We thank B. N. Taylor and P. Mohr for calling a possible problem
of sign to our attention. }
\eqna
  \Delta \nu (\mbox{weak})
   & =& - G_F { 3 \sqrt{2} m M \over  8 \alpha \pi } E_F
\nonumber \\
& \simeq &  -0.065 ~\mbox{kHz}.   \label{weak}
\endeqna
Numerical values of terms given by Eqs. (\ref{nonrecoil}) - ({\ref{weak}) 
are summarized in Table \ref{table0}.
If one uses the value of $\alpha$,
$R_{\infty}$ and $M/m$ from Refs. \cite{cage}, \cite{nez} and
\cite{mariam}:
\begin{eqnarray}
  \alpha^{-1}&=&137.035~997~9~(32)~~~(0.024~\rm{ppm})    \nonumber   \\
  R_{\infty} &=&10 ~973~~731.568~30 ~(31)~ {\rm m}^{-1},  \nonumber   \\
  {M   \over  m } &=& 206.768~259~(62),    \label{constants}
\end{eqnarray}
theoretical prediction for the hyperfine splitting of
the ground state muonium,
sum of the contributions listed in Table \ref{table0}, is given by
\eqn
\Delta\nu (\mbox{old theory})= 4~463~302.27~ (1.34)~(0.21)~(0.16)~(1.00) 
\label{oldtheory}
\endeqn
where the first and second errors reflect the uncertainties in the
measurements of $m_{\mu}$ and $\alpha^{-1}$ listed in (\ref{constants}).
The third error is purely theoretical and dominated by the uncertainty
in the last $\alpha (Z\alpha)^2$ term of (\ref{nonrecoil}).
The last one, about 1 kHz,  is an estimated contribution from
the order $\alpha^2(Z\alpha)$ correction in $\Delta({\rm rad})$. 
%[end of change.]
 
%%%
\begin{figure}[t]
\centerline{\epsfbox{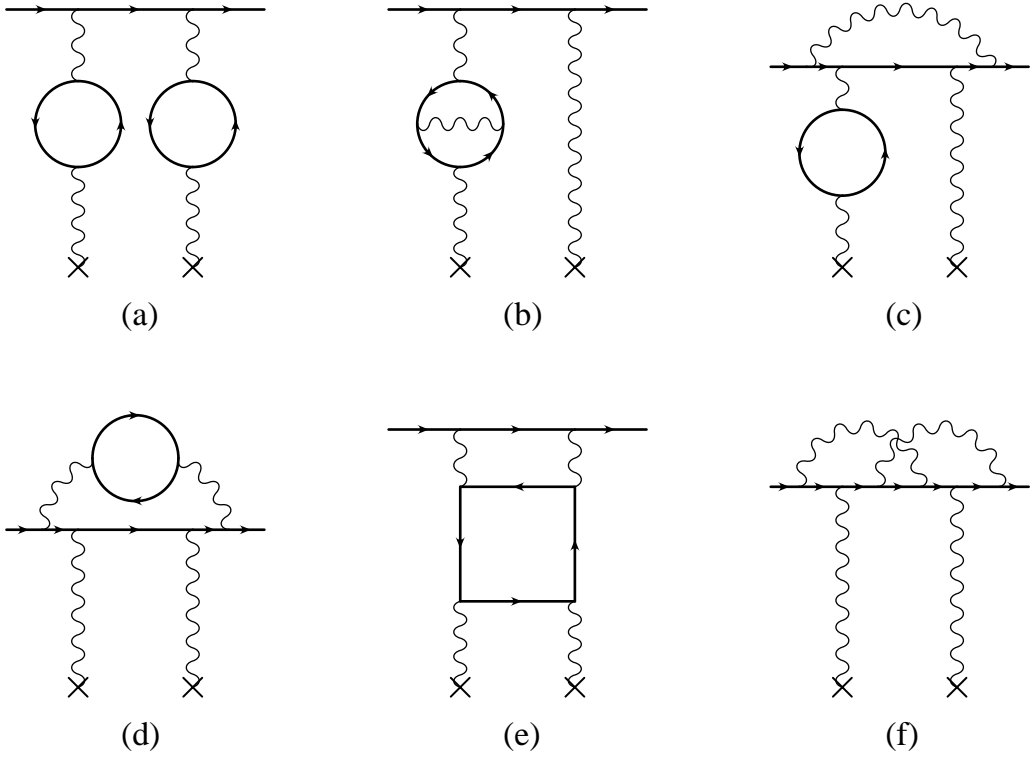}}
\vspace{3ex}
\caption{ Representative diagrams contributing to the
$\alpha^2 (Z \alpha)$ radiative corrections to the muonium hyperfine
structure in which two virtual photons are exchanged between
$e^-$ and $\mu^+$.
The muon is represented by $\times$.
\label{alldiagrams}}
\vspace{3ex}
\end{figure}
%%%%

As is clear from (\ref{oldtheory})
one must know the $\alpha^2 (Z\alpha)$ pure radiative correction
in order to improve the theoretical prediction further.
Fig. \ref{alldiagrams}  shows typical diagrams contributing to this order.
Recently, terms represented by the diagrams (a) - (e) of
Fig. \ref{alldiagrams} have been evaluated by Eides $et~al$. \cite{EKS2}.
Their results are as follows:
\eqna
  \Delta \nu (\mbox{Fig.1(a)}) &=& {36 \over 35} {{\alpha^2 (Z\alpha )}
 \over \pi} E_F  \nonumber  \\
&=& 0.567 ~3~\mbox{kHz} ,~~~~~~~~~~~~~~~~~~~~~~~~~~~~~~~~~ \label{fig1a}
\endeqna
\eqna
  \Delta \nu (\mbox{Fig.1(b)}) &=& \left ( {224 \over 15} \ln 2 - {38
\over 15}\pi -{118 \over 225} \right ) {{\alpha^2 (Z\alpha )}
\over \pi} E_F  \nonumber   \\
&=& 1.030 ~2~\mbox{kHz} ,~~~~~~~~~~~~~~~~~~~~~~~~~~~~~~~~   \label{fig1b}
\endeqna
\eqna
  \Delta \nu (\mbox{Fig.1(c)})&=& \left (
 -{4 \over 3} z^2  -{{20\sqrt{5}} \over 9} z - {64 \over 45} \ln 2
 + {\pi^2 \over 9} + {1043 \over 675} + {3 \over 8} \right )
{{\alpha^2 (Z\alpha )} \over \pi} E_F  \nonumber   \\
&=& -0.368 ~9~\mbox{kHz} ,   \label{fig1c}
\endeqna
\eqna
\Delta \nu (\mbox{Fig.1(d)}) &=& -0.310~742 \cdots ~{{\alpha^2 (Z\alpha )}
\over \pi} E_F  \nonumber   \\
&=& -0.171 ~4~\mbox{kHz},~~~~~~~~~~~~~~~~~~~~~~~~~~~~~~~~~~\label{fig1d}
\endeqna
where $z = \ln ((1 + \sqrt{5})/2)$.
The results (\ref{fig1a}), (\ref{fig1b}) and (\ref{fig1c}) are analytic,
while
(\ref{fig1d}) was evaluated numerically after reducing the integral
to one dimension.
We confirmed these results by an independent numerical calculation.
However, our purely numerical evaluation of Fig. 1(e):
\eqna
  \Delta \nu (\mbox{Fig.1(e)}) &=& -0.472~48~(9) {{\alpha^2 (Z\alpha )}
 \over \pi} E_F  \nonumber   \\
&=& -0.260 ~6~\mbox{kHz}  ~~~~~~~~~~~~~~~~~~~~~~~~~~~~~~~~~\label{ourfig1e}
\endeqna
disagreed with the semi-analytic result of Ref. \cite{EKS3}.
With our help, Eides \cite{Eides} found an error in the Table
after Eq. (23) of Ref. \cite{EKS3}.
Their corrected value is in good agreement with (\ref{ourfig1e}).

%%%
\begin{figure}[p]
\centerline{\epsfbox{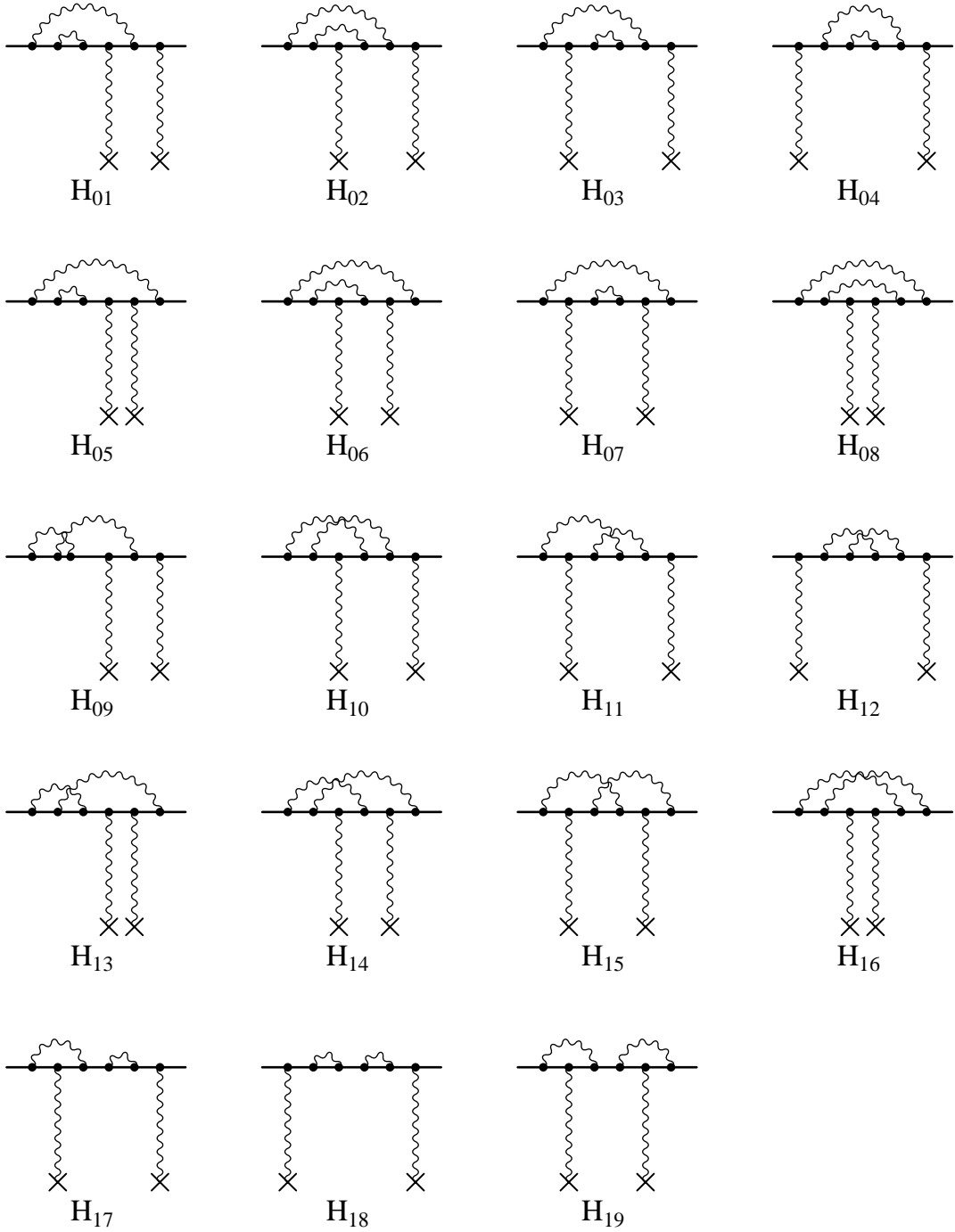}}
\vspace{3ex}
\caption{Two-photon exchange diagrams with fourth-order
radiative corrections on the electron-line.
Diagrams which are related to these diagrams by
time reversal are not shown explicitly.
The muon is represented by $\times$.
\label{ourdiagrams}}
\vspace{3ex}
\end{figure}
%%%

Fig. \ref{ourdiagrams} shows the complete set of
Feynman diagrams of the type (f) of Fig. \ref{alldiagrams},
which have not been evaluated before our work \cite{KN1}.
The preliminary result of our calculation for
all diagrams of Fig. \ref{ourdiagrams} was
\eqna
  \Delta \nu (\mbox{Fig.1(f)}) &=& -0.63~(4) {{\alpha^2 (Z\alpha )}
\over \pi} E_F
 \nonumber  \\
&=& -0.347~(22)~\mbox{kHz} , ~~~~~~~~~~~~~~~~~~~~\label{KN1result}
\endeqna
where the error is mainly due to the uncertainty in extrapolating
the integral to zero infrared cutoff.
The main purpose of this paper is to report a further improvement of this
result:
\eqna
  \Delta \nu (\mbox{Fig.1(f)}) &=& -0.676~4~(79) {{\alpha^2 (Z\alpha )}
\over \pi} E_F
 \nonumber  \\
&=& -0.373~1~(44)~\mbox{kHz} . ~~~~~~~~~~~~~~~~~~~~\label{newresult}
\endeqna
As a consequence of this result,
the total contribution of the $\alpha^2(Z\alpha)$ correction
to the muonium hyperfine splitting becomes
\eqna
\Delta \nu (\mbox{Fig.1})  &=& ~0.767~9~(79) {{\alpha^2 (Z\alpha )}
\over \pi} E_F
 \nonumber  \\
&=& ~0.423~5~(44)~\mbox{kHz} .  
~~~~~~~~~~~~~~~~~~\label{alpha^2(Zalpha)}
\endeqna
This removes the dominant theoretical uncertainty in 
$\Delta \nu(\mbox{theory})$.

In Sec. \ref{sec:NRQED} we  outline the NRQED treatment
of two-body bound system. It serves as the theoretical basis for
the calculation of the $\alpha^2(Z\alpha)$ correction as well as
the calculation of the $\alpha(Z\alpha)^2$  and  higher order corrections
discussed in the subsequent papers.
In Sec. \ref{sec:KP}  we illustrate 
the  general procedure of NRQED choosing  
the well-known $\alpha(Z\alpha)$ non-recoil radiative correction
as an  example. 
In Sec. \ref{sec:KN} we present our calculation of 
the $\alpha^{2}(Z\alpha)$ purely radiative
non-recoil correction to the muonium hyperfine structure.  Some problems
encountered in the numerical work are also discussed there.
Sec. \ref{sec:DISCUSS} is devoted to the discussion of our results.

%%%% Sec.2
\section{NRQED}
\label{sec:NRQED}
\subsection{Why NRQED ?}
\label{subsec:why}

The Lorentz  invariance has been one of the most important
guiding principles for the development of quantum field theory.
However,  relativistic quantum field theory
is often very cumbersome
to apply to nonrelativistic bound systems.
Such a calculation tends to be very complicated and requires
an enormous effort, while the result reflects mostly the
nonrelativistic feature of the system.
For such a system an approach that incorporates most of
the bound state effects from the beginning
would minimize
the amount of computation necessary to
achieve the desired precision.
For the case of electromagnetic interaction,
this has been realized by a theory called 
nonrelativistic quantum electrodynamics, or NRQED.
The NRQED enables us to avoid some, if not all, of the problems
encountered in the usual treatment based on the Bethe-Salpeter equation.

The NRQED, formulated by Caswell and Lepage \cite{CL}, is
a rigorous adaptation of QED to bound systems.
This theory enables us to take a consistent and systematic approach to 
loosely bound  nonrelativistic systems.
Compared with   the conventional bound state theories,
it allows easier power counting,
more transparent cancellation of UV and IR divergences,
and is manifestly gauge invariant.
In spite of its superiority,
however, the details of the theory
has not yet been fully worked out.
In this series of papers, we present an explicit construction
of the NRQED Hamiltonian and develop a bound state
perturbation theory based on it.

As for  the computation of the $\alpha(Z\alpha)$ and $\alpha^2(Z\alpha)$
corrections,  NRQED or any other relativistic bound state formalism  
gives the same simple recipe: calculate the forward scattering amplitude 
in QED and multiply it with 
$|\phi(0)|^2$, where $\phi(0)$ is the nonrelativistic wave function
at the origin. In NRQED this recipe can be  directly justified 
by inspection of relevant diagrams and power counting.
In other bound state formalisms, 
the corresponding procedure may be  less straightforward.
The latter approach becomes very complicated
for higher-order corrections such as  the $\alpha(Z\alpha)^2$ and 
$\alpha(Z\alpha)^3$ corrections.  
Difficulty in achieving high numerical
precision by this method is one of the  sources
of theoretical uncertainty at present \cite{SY,sapirstein}.

The approach adopted by the NRQED, however, loses its effectiveness for
the high Z system. In such a case it is desirable to avoid expanding in
$v \sim Z\alpha$.
Recently,  an attempt has  been made to
calculate the order $\alpha$  term without expanding in $Z\alpha$  
\cite{persson}. 
However,  this approach may have difficulty in providing a good 
precision for $Z=1$.
This is primarily because, for  low $Z$ systems, the bound electron is
almost  on-the-mass-shell and causes the near infrared divergence. 
As a consequence the convergence of   numerical integration
deteriorates as $Z$ decreases.  
As is shown in the  subsequent papers, 
the NRQED method enables us to  deal with the near infrared divergence 
problem order by order in a
systematic expansion in $Z\alpha$, and allows us to
calculate the expansion coefficients with high precision.
This is why the NRQED method is a powerful tool for  low $Z$ systems.

\subsection{Outline of NRQED }
\label{subsec:outline}

In the NRQED approach to the bound state problem, one
first derives the NRQED Lagrangian  from  the QED, and then
uses it to determine the correction to the energies and wave functions
by a systematic application of the Rayleigh-Schr\"{o}dinger
perturbation theory.

The NRQED Lagrangian consists of  all  possible local
interactions satisfying the required symmetries, such as
gauge invariance, parity invariance, time reversal,
galileian invariance, hermiticity,
and locality.
We use the same  photon Lagrangian $(-1/4 )F_{\mu \nu }F^{\mu \nu}$ 
as that of QED.
In addition, new photon interaction terms are introduced to 
represent the insertion of the fermion loop, such as vacuum polarization
and light-by-light scattering.  

In order to define the NRQED Lagrangian precisely, we must
regularize the interaction terms of NRQED,
e.g., by cutting off contributions of large momenta.
Since this theory is meant to apply to nonrelativistic systems,
the cut-off $\Lambda$ may be chosen as the typical mass
scale of the system, e.g.,  the rest mass of an electron.
With the cut-off $\Lambda$ thus fixed, the theory becomes well-defined,
even though the interaction terms are strongly dependent on the cut-off
parameter.
In the following the cut-off is understood implicitly,
and will be exhibited only when it is necessary.
The choice of the momentum cut-off used for the
NRQED scattering amplitudes is  arbitrary but the  physical
quantity computed should be independent of any particular choice.
In other words, the NRQED theory must have reparametrization
invariance with respect to the choice of cut-off.
This is analogous to the existence of the renormalization
group in the renormalizable relativistic field theory.
It is important to note that the NRQED is  fully equivalent to the QED.
The only difference is that it is better adapted to low energy
bound systems.

The NRQED rule for determining
the operators which appear in its Lagrangian
and their coefficients is simple and straightforward:
Each term of the $scattering$ amplitude calculated in the  NRQED
must coincide with the corresponding $scattering$ amplitude
of the original QED at some given momentum scale,
e.g., at the threshold of the external on-shell particles.
The center of  mass frame is used
for  both bound state and  scattering state calculations.
Since the same argument about reparametrization invariance
holds for the momentum scale chosen for comparison
of QED and NRQED scattering amplitudes,  the  at-threshold
condition is just for convenience.
However,  the on-shell condition for the external fermion is
more than a matter of convenience.
In order to regulate the IR singularity
it is convenient to introduce the photon mass $\lambda$ in
the calculation of scattering amplitude of both QED and NRQED.
This finite photon mass together with the on-shell condition
ensures that the NRQED scattering theory  has a pole in the region of
the complex  energy plane of the external fermion
in which  the scattering theory  can be  analytically continued to
the off-shell bound state theory.

We use the  normalization $u^{\dag}u=1$
for  the external 4-component spinors in the QED calculation
instead of the conventional relativistic
normalization $\bar{u}u=1$ so that both QED and NRQED
$S$-matrix  have the same normalization \cite{nrqcd}.
This ensures that  physical quantities,
such as  decay rate and  cross section,
calculated in both theories are the same.

Note that the scattering amplitude of QED is fully renormalized,
namely, it is finite and completely determined within QED.
This enables us to fix the NRQED $``$renormalization" constants without
ambiguity.
This also means that the coupling constant $\alpha$  and
fermion  masses in the QED are
the renormalized ones determined on-shell, and these $\alpha$
and fermion masses are used
as the $``$bare" coupling constant and $``$bare" masses of NRQED.

It is convenient to write the NRQED Lagrangian in two parts:
$L_{\rm main}$ and $L_{\rm contact}$.
The $L_{\rm main}$ part consists of the fermion bilinear operators.
Fermions in NRQED are expressed by the Pauli two component
spinor field $\psi(t,\vec{x})$ (instead of the Dirac spinor).
If one takes  into account the  required symmetries of the theory,
the main part of NRQED Lagrangian $L_{\rm main}$  must have the general form
\cite{CL,KL}
\eqna
L_{{\rm main}}& = &\psi^{\dagger}\{ iD_{t} + \frac{\vec{D}^{2}}{2m}
+ \frac{\vec{D}^4}{8m^{3}}
 \nonumber \\
& &
+ c_{F}\frac{e \vec{\sigma}\cdot\vec{B}}{2m}
+ c_{D}\frac{e (\vec{D}\cdot\vec{E} -\vec{E}\cdot\vec{D})}{8m^{2}}
\nonumber \\
& &
+ c_{S}\frac{ie \vec{\sigma}\cdot(\vec{D}\times \vec{E}
                                - \vec{E}\times \vec{D} )}{8m^{2}}
+ c_{W1}\frac{e \{\vec{D}^{2},\vec{\sigma}\cdot\vec{B}\}}{8m^{3}}
\nonumber \\
& &
+c_{W2}\frac{-e \vec{D}^i\vec{\sigma}\cdot\vec{B}\vec{D}^i}{4m^{3}}
+c_{p'p}\frac{e (\vec{\sigma}\cdot\vec{D}\vec{B}\cdot\vec{D}
                 +\vec{D}\cdot\vec{B}\vec{\sigma}\cdot\vec{D}) }
            {8m^{3}}
+ \dots \}\psi ~.      \label{Lagra}
\endeqna
where $D_{t}=\partial_{t}+ieA^0 $
and $\vec{D}=\vec{\partial}-ie\vec{A}$.
(We put  $ c=1 $ and $\hbar=1 $ henceforth.)
The positron  part can be written down in a similar way.
The particle-antiparticle mixed interaction is not present
in $L_{\rm main}$.
The first three terms are related to the kinetic term of the
QED Lagrangian.
The second and third terms are derived from the  expansion
\eqn
E=\sqrt{\vec{p}\,^2+m^2}=m+{\vec{p}\,^2 \over 2m}
-{\vec{p}\,^4 \over 8m^3}+~\ldots~~ .
\endeqn
These three terms of (\ref{Lagra})
have  coefficients unaffected by the radiative correction
as a consequence of the renormalizability of  QED,
while the coefficients $c_i$ of other terms are
modified by the QED interaction
and can be expressed as a power series in the coupling
constants $\alpha$
\eqn
c_i=c_i^{(0)} +  c_i^{(1)} \alpha + c_i^{(2) }\alpha^2 + \ldots~~~~.
\label{coeffC}
\endeqn
Some of the  operators in (\ref{Lagra}) can be  generated by the
Foldy-Wouthuysen-Tani transformation of
the Dirac Lagrangian. These operators have the coefficient $c_i^{(0)}=1$
while other operators have $c_i^{(0)}=0$.
Note that $c_i$'s do not have coefficients involving $Z\alpha$
caused by the binding effect
because they are determined solely by comparison of the NRQED and
QED {\it scattering} amplitudes without  referring to the bound states.

Eq. (\ref{Lagra}) has an infinite number
of terms.  Not all of them, of course,  are needed in
a practical calculation.
The operators necessary to carry out a particular calculation
are determined by the power counting rule of NRQED
for the bound state. We will show this process explicitly
in the next subsections where the NRQED Hamiltonian is constructed.

The $L_{\rm main}$ of NRQED alone is not sufficient to  produce 
the same physical quantities as those from QED. 
To make NRQED equivalent to QED,  
we must add  another term to the  NRQED Lagrangian.
%The second part of NRQED Lagrangian $L_{\rm contact}$
It consists of terms of contact interaction type:
\eqna
 L_{\rm contact}& = &
 d_1 {1 \over mM } (\psi^{\dagger} \vec{\sigma }\psi)
 \cdot (\chi^{\dagger} \vec{\sigma }\chi)
+d_2 {1 \over mM } (\psi^{\dagger} \psi) (\chi^{\dagger} \chi)
\nonumber \\
& + & d_3 {1 \over mM } (\psi^{\dagger} \vec{\sigma }\chi)
\cdot (\chi^{\dagger} \vec{\sigma }\psi)
     +d_4 {1 \over mM } (\psi^{\dagger} \chi )(\chi^{\dagger} \psi)
\nonumber \\
& + & d_5 {1 \over  m^3M } (\psi^{\dagger} \vec{D}^2\vec{\sigma }\psi)
\cdot  (\chi^{\dagger} \vec{\sigma }\chi) +~\cdots~~   ,   \label{contact}
\endeqna
where $\chi$ represents a fermion field (of mass $M$) such as
a muon or a positron.
The third and fourth terms in (\ref{contact}) are needed only
when both electron and positron are present.
This is because, from the viewpoint of NRQED,
the electron-positron annihilation 
is a high energy process and can only be represented
as a contact interaction term.
For the muonium, only the first and second terms are relevant.
The fifth term is an example of contact terms including derivative
interactions, which are of  higher order
in $< \vec{p}\,^2/m^2 > \sim (Z\alpha)^2 $.

The coefficients $d_i$ are chosen such that 
these contact interactions
make up the  difference between the  QED electron-muon  scattering 
amplitude and the corresponding NRQED  scattering
amplitude derived from the  Lagrangian $L_{\rm main}$.
This procedure enables us to determine the
coefficients $d_i$ completely.

As is clear from the above discussion, these NRQED $``$renormalization"
constants $c_i$ and
$d_i$ have the  parameter dependence
\eqna
&& c_i = c_i ( \alpha, \Lambda, m ),
\nonumber \\
&& d_i = d_i ( \alpha, Z\alpha, Z^2\alpha,\Lambda, m, M ). \label{Cdpara}
\endeqna
Of course, the experimentally observable  result of calculation
must be independent
of the cut-off $\Lambda$,
and gauge invariant.  This is realized by a systematic application
of the nonrelativistic Rayleigh-Schr\"{o}dinger
perturbation theory to the bound states.
Note also that $c_i$ and $d_i$ are finite and well-defined
in the infrared limit and hence require no infrared cut-off.

Just as the actual execution of renormalization program of
QED must rely on the covariant perturbation theory,
a comprehensive formulation of NRQED
can be realized explicitly only within the framework of
the nonrelativistic Rayleigh-Schr\"{o}dinger perturbation theory.
This means that we have to choose an appropriate part of the Hamiltonian
as the unperturbed term and treat the rest as perturbation.

To deal with the muonium we find it generally convenient to define
the unperturbed system
in terms of the ground state solution of the nonrelativistic
Schr\"{o}dinger equation:
\eqn
\left(\frac{\vec{p}\,^{2}}{2m_{r}}-\frac{Z\alpha}{r}\right)
\phi= E^{0} \phi ~,
\endeqn
where $m_{r}$ is the reduced mass,  $E^{0}=-\gamma^2/(2m_r)$
is the ground state binding energy,
$\gamma \equiv (Z\alpha) m_{r}$ being a typical momentum scale of
the Coulomb bound state.
The solution of this equation is
\eqn
\phi(\vec{p})= \frac{8\sqrt{\pi\gamma^{5}}}
{(\vec{p}\,^2+\gamma^{2})^{2}}~.
\endeqn
The unperturbed electron field $\psi(\vec{p})$ is thus expressed
by this wave function $\phi(\vec{p})$ times the Pauli spin factor.
Using the  remaining interaction terms in the NRQED Lagrangian
together with the photon Lagrangian, we can construct the effective
potentials.
These potentials are to be treated as perturbation.

The nonrelativistic Rayleigh-Schr\"{o}dinger perturbation theory
gives
\eqna
  \Delta E_{n} & = &  \psi^{\dagger }_{n} V \psi_{n}
         [
         1+  \psi^{\dagger }_{n} \left(\frac{\partial}{\partial E}
         V \right) \psi_{n}   ] _{E=E_{n}^{0}}
\nonumber \\
               & + &  \psi^{\dagger }_{n} V
                   (\tilde{G}_{0}-\frac{\psi_{n}\psi^{\dagger}_{n}}
                   {E-E_{n}^{0}})
         V \psi_{n}  ~|_{E=E_{n}^{0}}
\nonumber \\
               & + & \psi^{\dagger }_{n} V
                   (\tilde{G}_{0}-\frac{\psi_{n}\psi^{\dagger}_{n}}
                   {E-E_{n}^{0}}) V
                   (\tilde{G}_{0}-\frac{\psi_{n}\psi^{\dagger}_{n}}
                   {E-E_{n}^{0}}) V \psi_{n}  ~|_{E=E_{n}^{0}}
\nonumber \\
               & + &  \psi^{\dagger }_{n} V \psi_n
               \psi_n^{\dagger} \left( \frac{\partial}{\partial E}
       \left(V (\tilde{G}_{0}-\frac{\psi_{n}\psi^{\dagger}_{n}} {E-E_{n}^{0}})
                    V \right) \right) \psi_{n} ~|_{E=E_{n}^{0}}
\nonumber \\
               & + &   \ldots~~ .  \label{perturb}
\endeqna
The Green function $\tilde{G}_0(\vec{k},\vec{q}:E_n^0)$
appearing here is known in a closed form
for the nonrelativistic Coulomb potential
\cite{schwinger}. For the ground state $n=1$, we find 
\eqna
\lim_{E\rightarrow E^{0}_{n=1}}(
\tilde{G}_{0}-\frac{\psi_{n=1}\psi^{\dagger}_{n=1}}
{E-E_{n=1}^{0}})
 & = &
\frac{-2m_r}{\vec{k}^2+\gamma^2}(2\pi)^3\delta^3(\vec{k}-\vec{q})
\nonumber  \\
 & + &
\frac{-2m_r}{(\vec{k}\,^2+\gamma^2)}
\frac{-Ze^2}{|\vec{k}-\vec{q}|^2}
\frac{-2m_r}{(\vec{q}\,^2+\gamma^2)}
\nonumber  \\
 & - & \frac{64\pi}{Z\alpha\gamma^4}\tilde{R}(\vec{k},\vec{q})~,
                        \label{green}
\endeqna
where
\eqna
\tilde{R}(\vec{k},\vec{q})  = && \frac{\gamma^{8}}
{(\vec{k}\,^{2}+\gamma^{2})^{2}
(\vec{q}\,^{2}+\gamma^{2}) ^{2}}
       \biggl[  ~\frac{5}{2}
- 4\frac{\gamma^{2}}{\vec{k}\,^{2}+\gamma^{2}}
- 4\frac{\gamma^{2}}{\vec{q}\,^{2}+\gamma^{2}}
\nonumber \\
&&+\frac{1}{2} \log A
          + \frac{2A-1}{(4A-1)^{1/2}}\tan^{-1}(4A-1)^{1/2} \biggr ]~,
\endeqna
and
\eqn
A=\frac{(\vec{k}\,^{2}+\gamma^{2})(\vec{q}\,^{2}+\gamma^{2})}{4\gamma^{2}
|\vec{k}-\vec{q}|^{2}}~.
\endeqn
The first, second, and third terms of
the expression (\ref{green}) can be understood as corresponding to
zero, one, and two or more Coulomb-photon exchanges.

In order to determine which terms of the Hamiltonian are needed to
obtain the desired
precision it is useful to know the expectation values of various
operators with respect to appropriate wave functions \cite{nrqcd}.
For a nonrelativistic Coulombic bound system, one finds
\eqna
&& < \vec{\partial}> \sim  m(v/c), ~~~
< \partial_t> \sim  m(v/c)^{2},~~~
< eA^0>  \sim  m(v/c)^{2},
 \nonumber \\
&&<e\vec{A}>  \sim  m(v/c)^{3}, ~~
<e\vec{E}>  \sim m^{2}(v/c)^{3}, ~~
<e\vec{B}>  \sim m^{2}(v/c)^{4},   \label{order}
\endeqna
where $ m $ is the  electron mass, $ v $ is the typical velocity
of a bound electron, and $ c(=1) $ is the velocity of light.
Thus, in Eq. (\ref{Lagra}), the first two terms,
next four terms, and the remaining terms
correspond to  the interactions which start at  
orders  $v^{2}, v^{4}$, and $v^{6}$,
respectively.
Radiative corrections, which alter the values of
the coefficients $c_i$'s and $d_i$'s,
will keep the estimate (\ref{order}) intact.
The information (\ref{order}) can be used to terminate
the series of interaction terms at the desired precision.
In this sense, the NRQED Lagrangian is an
expansion in both the coupling constant $\alpha$
and the velocity $v$.

The NRQED $``$renormalization" coefficients
play important roles in restoring
gauge invariance which might have been broken by regularization.
The explicit form of these coefficients depends on the regularization
method.  Gauge invariant regularization is desirable but not necessary.
If one proceeds carefully, even a simple momentum cut-off method
may be used \cite{zwanziger}.
(This is only true for an Abelian  gauge theory
such as NRQED.)
In a calculation of the would-be divergent quantity
in NRQED, we  put the UV cut-off $\Lambda$
not in the fermion momentum but in the photon momentum \cite{pat}.

Because of the way NRQED is constructed,
gauge invariance of  the NRQED amplitude with its complete
set of $``$renormalization" constants is automatically
guaranteed by the gauge invariance of the corresponding QED amplitude.
Since QED and NRQED are separately
gauge invariant, we may choose different gauges
for QED and NRQED. We will use the Feynman gauge for
QED calculation, and we use  the Coulomb gauge  for NRQED.
The Feynman gauge minimizes the amount of work for numerical computation,
and the Coulomb gauge is more suitable for describing the
nonrelativistic behavior of the electron.

The dominant contribution to the hyperfine splitting between
the spin J=1 and J=0 states
originates from the interaction between
the electron spin and the muon spin mediated by a transverse photon
of momentum $\vec{k}$.  This leads to a potential of the form
\eqn
 V_{F} = \frac{ie}{2m}(\psi^{\dagger}\vec{k}\times
\vec{\sigma_{e}}\psi )
        \cdot
       \frac{-iZe}{2M}(\chi^{\dagger}
       (-\vec{k})\times\vec{\sigma_{\mu}}\chi)
       \frac{-1}{\vec{k}\,^2}   \label{Fpotential}
\endeqn
in the momentum space representation, where $ M $ is the muon mass.
Using this Fermi potential $V_F$ in  the first order
perturbation theory and taking the difference
between J=1 and J=0 states, we obtain
the hyperfine Fermi splitting
\eqna
 E_{F}& = & \langle n=1| V_{F} | n=1 \rangle |^{J=1}_{J=0}
\nonumber \\
     &  = & \int{d^3p \over (2\pi)^3}\int{d^3k \over (2\pi)^3 }
 \frac{(8\sqrt{\pi\gamma^{5}})^2}
{(\vec{p}\,^{2}+\gamma^{2})^{2} (|\vec{p}+\vec{k}|^{2}+\gamma^{2})^{2}}.
\nonumber \\
    & & \frac{ie}{2m}\langle(\vec{k}\times\vec{\sigma_{e}})\cdot
       \frac{-iZe}{2M}((-\vec{k})\times\vec{\sigma_{\mu}})\rangle
       \frac{-1}{\vec{k}\,^2}
\nonumber \\
    & = &{ 2(Z\alpha)\gamma^3 \over 3mM }
        \langle \vec{\sigma_e}\cdot\vec{\sigma_\mu}\rangle|^{J=1}_{J=0}
\nonumber \\
    & = & {8(Z\alpha)\gamma^3 \over 3~mM}~.    \label{E_F}
\endeqna
Needless to say, the Fermi potential is of order $v^4 (m/M)m
\sim (Z\alpha)^4(m/M) m$.

\begin{figure}[p]
\vspace{-3ex}
\centerline{\epsfbox{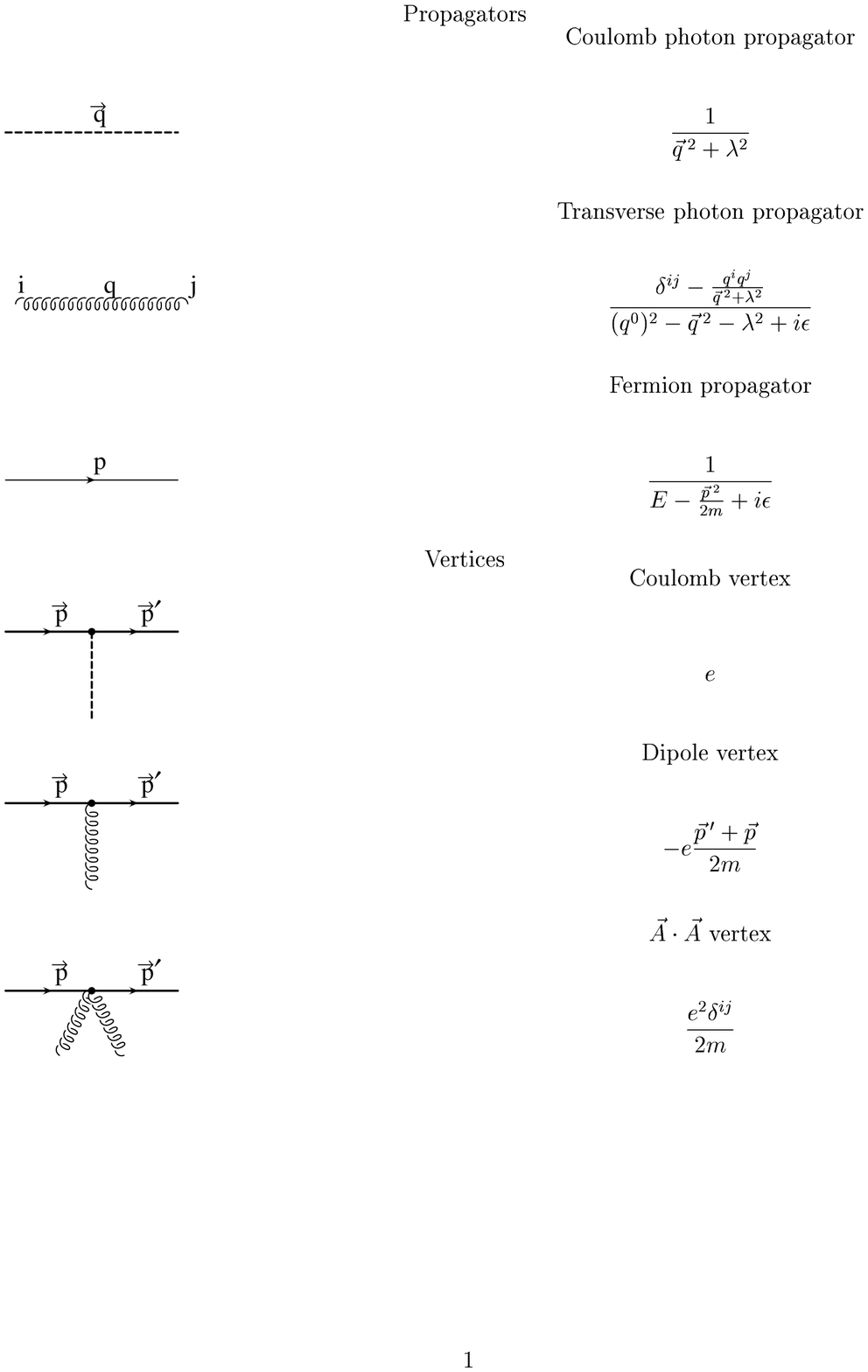}}
\caption{ NRQED $``$Feynman" rules for vertices and propagators.
They can be used for both scattering and
bound state calculations.
$E$ in  the fermion propagator represents the fermion's bound state
energy: $E=0$ for scattering and $E=-\gamma^2/(2m_r)$ for the ground state
muonium.  The photon mass $\lambda$ is set to  zero in
the bound state calculation.
\label{Feynrule}}
\end{figure}
%%%%
\begin{figure}[p]
\centerline{Figure 3 (continued)}
\centerline{\epsfbox{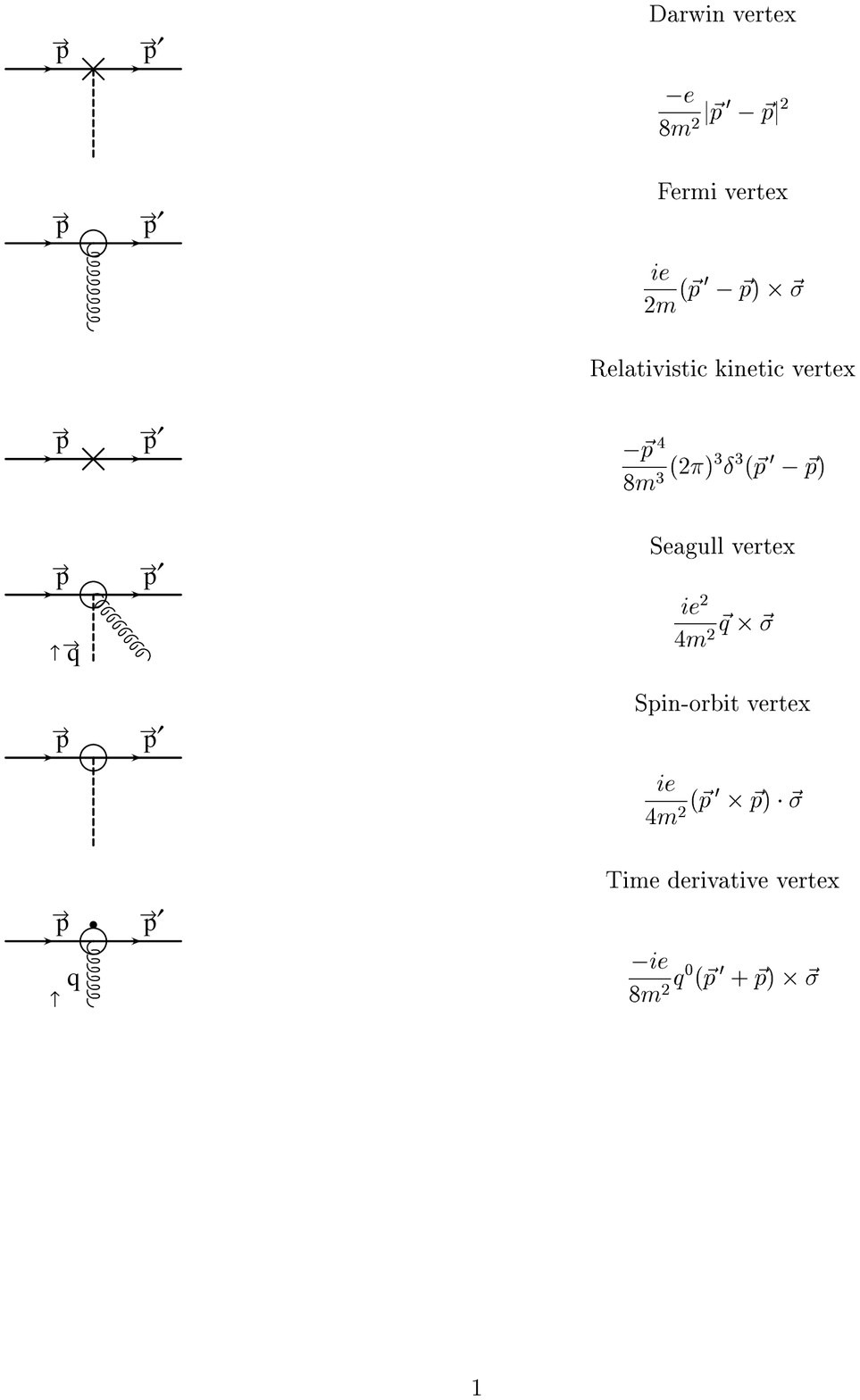}}
\end{figure}
%%%%
%%%%
\begin{figure}[t]
\centerline{Figure 3 (continued)}
\centerline{\epsfbox{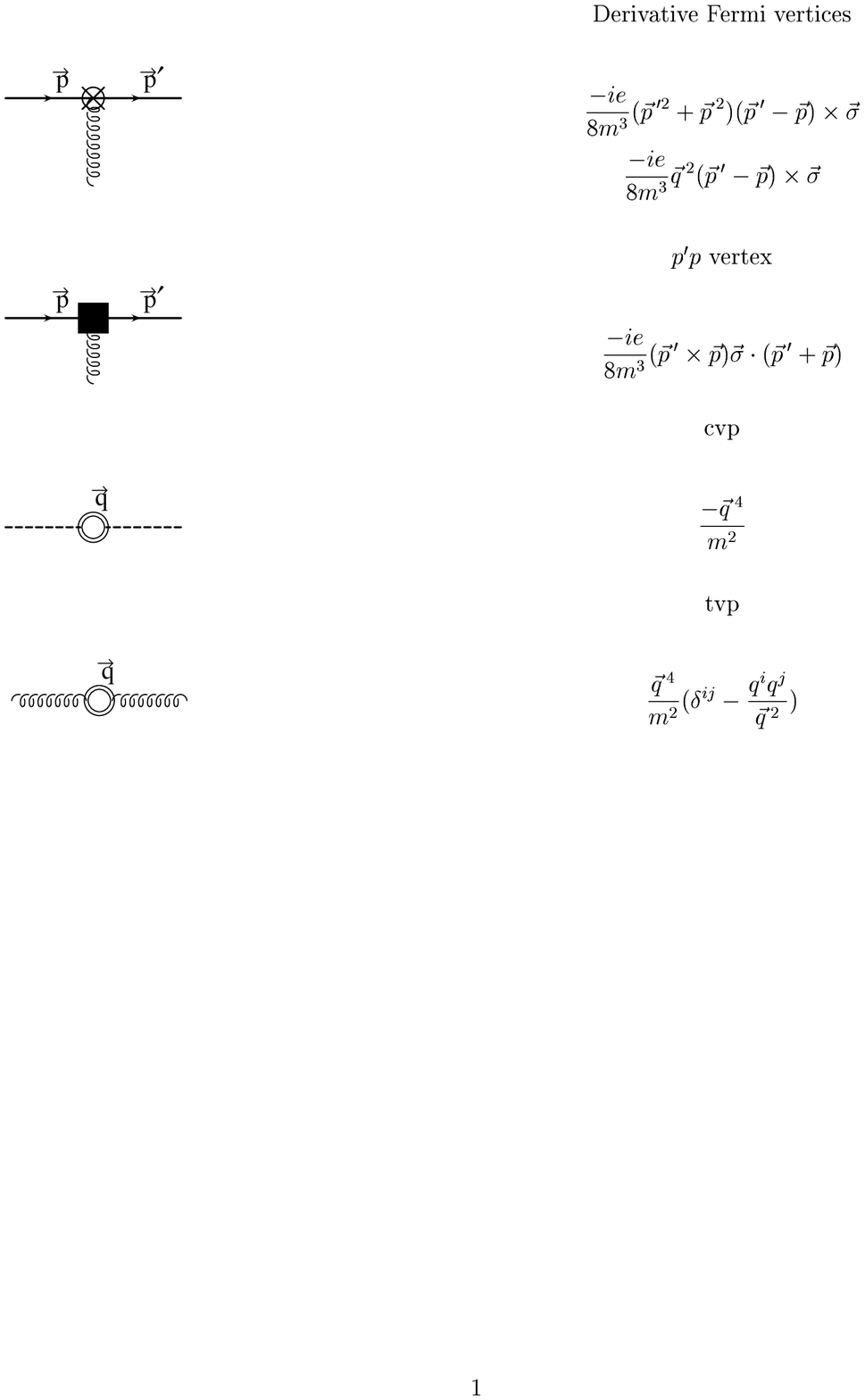}}
\end{figure}
%%%%

Various interaction terms and propagators are represented by
the NRQED $``$Feynman" diagrams shown in Fig. \ref{Feynrule}.
It is convenient and useful to express higher-order amplitudes
representing scattering states or bound states by corresponding
diagrams.

We classify the diagrams according
to the number of external  photons in the QED Feynman diagrams.
In the following subsections
we shall show step by step how the corresponding NRQED Lagrangian
$L_{\rm main}$ (or Hamiltonian $H_{\rm main}$) is determined.

\subsection{Scattering by a static external potential}
\label{subsec:one-photon}

We have already found the general form of the main part of
the NRQED Lagrangian given by Eq. (\ref{Lagra}) using the required symmetries 
for the theory and the power counting rules. 
Therefore, the remaining task for  construction of the NRQED Hamiltonian 
is  determination of the coefficients of these operators
appearing in Eq. (\ref{Lagra}).

Let us first consider the QED diagram in which one photon is
exchanged between the electron and the muon.
The first step to obtain the $``$renormalization" coefficients  
of the operators in the NRQED Hamiltonian
is to carry out    nonrelativistic reduction of
the QED scattering amplitude exchanging one photon between
the electron and the muon.
Comparing this  QED scattering amplitude with the scattering amplitude 
derived from the general form of NRQED Lagrangian given 
in Eq.  (\ref{Lagra}), we are able to fix the $``$renormalization"
coefficient $c_i$'s.
We want to chose the simplest  process to find them. 
It turns out that all $``$renormalization" coefficients 
in (\ref{Lagra}) can be
obtained by using  the external static potential. 
The  comparison between the corresponding QED and NRQED amplitudes
is shown in Fig. \ref{comparison1}.
%
%
%
%%%%%
\begin{figure}[p]
\centerline{\epsfbox{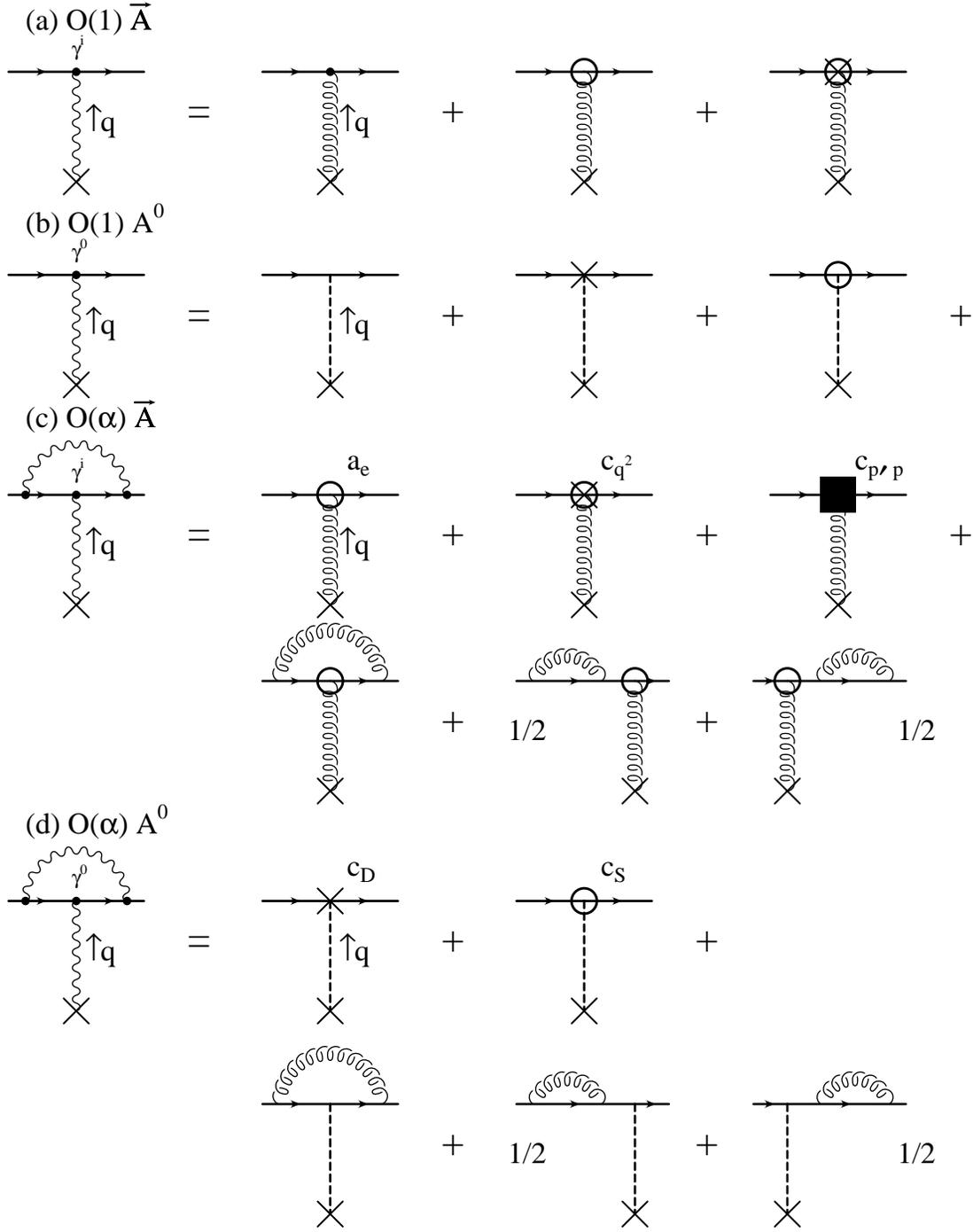}}
\vspace{3ex}
\caption{ QED and NRQED scattering diagram comparison.
The diagrams on the left and  right of the = sign represents
QED and NRQED diagrams, respectively.
The external fermions are on-the-mass-shell and at threshold.
Self-energy diagrams coming from the $\vec{A}\cdot\vec{A}\psi^{\dagger}\psi$
vertex as well as self-mass counterterms are not shown explicitly.
\label{comparison1}}
\vspace{3ex}
\end{figure}
We will work out nonrelativistic reduction of operators of order up to
$\alpha v^6$ and $\alpha v^4$  for spin-flip and  spin-non-flip
ones, respectively. Since the spin-non-flip operators contribute to the
hyperfine splitting only through the higher order bound state perturbation,
we need only operators of order lower than the spin-flip ones.
The QED scattering amplitude to be studied here consists of  a tree vertex and
one  dressed by a radiative photon.
Because we are dealing with the scattering amplitude, we also have
a diagram with self-energy insertion on the external fermion lines.
However, these diagrams can be dropped after the
mass renormalization and  wave function renormalization
are carried out, if one chooses the on-shell renormalization scheme.
Then the QED scattering amplitude is expressed by the usual
form factors $F_1$ and $F_2$.

For the external static vector potential $\vec{A}(\vec{q})$, we easily find
the QED scattering amplitude
\eqna
&&e\bar{u}(\vec{p}\,') \biggl [
-\vec{\gamma}\cdot\vec{A}(\vec{q})
     F_1(q^2)
   +{i \over 2m} \sigma^{ij}
    A^i(\vec{q})q^jF_2(q^2)~ \biggr ] ~u(\vec{p})
\nonumber \\
& & =  F_1(q^2) \psi^{\dagger} (\vec{p}\,')
\biggl [ ~ -{ e \over 2m} (\vec{p}\,'+\vec{p})\cdot\vec{A}
- {ie \over 2m} \vec{\sigma}\cdot(\vec{q}\times\vec{A})
\nonumber \\
&&+{ie \over 8m^3}(\vec{p}\,'^2+\vec{p}\,^2)
\vec{\sigma}\cdot(\vec{q}\times\vec{A}) +\ldots~~\biggr ]  \psi(\vec{p})
\nonumber \\
& &+ F_2(q^2)\psi^{\dagger}(\vec{p}\,')
\biggl [~ -{ ie \over 2m} \vec{\sigma}\cdot(\vec{q}\times\vec{A})
+{ie \over 16m^3}(\vec{p}\,'^2+\vec{p}\,^2)
\vec{\sigma}\cdot(\vec{q}\times\vec{A})
\nonumber \\
&&+ {ie \over 8m^3}\vec{\sigma}\cdot\vec{p}\,'
\vec{\sigma}\cdot(\vec{q}\times\vec{A})
\vec{\sigma}\cdot\vec{p}
 +\ldots~~ \biggr ]  \psi(\vec{p}),            \label{reduction1}
\endeqna
where $u$ and $\psi$ are Dirac and Pauli spinors, respectively.
Similarly, for the external static Coulomb field $A^0(\vec{q})$, we have
\eqna
&&e\bar{u}(\vec{p}\,')\biggl [\gamma^0 A^0(\vec{q}) F_1(q^2)
-{i \over 2m} \sigma^{0j}
A^0(\vec{q})q^{j}F_2(q^2)~\biggr ] ~u(\vec{p})
\nonumber \\
&& =  F_1(q^2)\psi^{\dagger}(\vec{p}\,')
\biggl [~ eA^0
- {e \over 8m^2}\vec{q}\,^2 A^0
+ {ie \over 4m^2}  \vec{\sigma}\cdot(\vec{p}\,'\times\vec{p})A^0
+ \ldots~~\biggr ] \psi(\vec{p})
\nonumber \\
&&+  F_2(q^2)\psi^{\dagger}(\vec{p}\,') \biggl [
{}~ -{e \over 4m^2}
\vec{q}\,^2 A^0
  + { i e \over 2m^2} \vec{\sigma}\cdot(\vec{p}\,'\times\vec{p})A^0
+ \ldots~~\biggr ]  \psi(\vec{p})~.           
\label{reduction2}
\endeqna

Taking account of the fact that $q^0$ is of order $v^2$
and $|\vec{q}|$ is of order $v$,   
the nonrelativistic expansion of the form factors
can be written as  \cite{Fey}
\eqna
F_1(q^2)&=& 1 - {\alpha \over 3 \pi}\biggl [ {\vec{q}\,^2 \over m^2}
                    \biggl( \ln \left({m \over \lambda }\right) 
                       - { 3 \over 8} \biggr ) \biggr ]
                   +{\cal{O}}(\alpha v^4, \alpha^2 v^2)
\nonumber \\
F_2(q^2)&=& a_e - {\alpha \over  \pi}{\vec{q}\,^2 \over 12 m^2}
  +{\cal{O}}( \alpha v^4,\alpha^2v^2)  , \label{formfactor}
\endeqna
where $a_e =F_2(0)$ is the anomalous magnetic moment
of the electron.

Combining  (\ref{reduction1}),(\ref{reduction2})
and (\ref{formfactor}) together,  and comparing with the scattering 
amplitude derived from the NRQED Lagrangian given by (\ref{Lagra}),
%we can construct the  main part of the NRQED Hamiltonian
%including their coefficients:
we find that the $``$renormalization" coefficients must be chosen as  
\eqna
c_F^{QED}& =&1+ a_e,
\nonumber \\
c_D^{QED}&=&1 + { \alpha \over \pi } {8 \over 3}
              \biggl [ \ln\left({m \over \lambda }\right)
    -{3 \over 8} \biggr ]
                 + 2 a_e,
\nonumber \\
c_S^{QED}&=& 1+ 2a_e ,
\nonumber \\
c_{W1}^{QED}&=& 1+  {\alpha\over\pi} {4 \over 3}\biggl [
\ln\left({m \over \lambda}\right)
- { 3 \over 8 }  +{1 \over 4} \biggr ]  + {a_e \over 2} ,
\nonumber \\
c_{W2}^{QED}&=& {\alpha\over\pi} {4 \over 3}\biggl [
\ln\left({m \over \lambda}\right)
- { 3 \over 8 }  +{1 \over 4} \biggr ]  + {a_e \over 2} ,
\nonumber \\
c_{p'p}^{QED}&=& a_e.
\endeqna

This procedure enables us to construct the NRQED Hamiltonian. 
However, it does not provide   the complete NRQED Hamiltonian.
One must also include  terms which are the NRQED analogues 
of QED counterterms such 
as $-\delta m \bar{u} u$. To see this let us calculate 
the NRQED scattering amplitude which  arises from the  Coulomb term
$ \psi^{\dagger} e A^0  \psi$ 
modified by a 1-loop NRQED radiative correction
by a perturbative treatment  of  $H_{\rm main}$.

The  perturbation here means that
the zeroth-order of NRQED Hamiltonian contains only the free part
of the electron, and thus  the Coulomb interaction is treated as
perturbation.
Some of these scattering amplitudes involving radiative
corrections  require new forms of the NRQED operators
while others may be represented by additional $``$renormalization"
constants of the already existing operators in $H_{\rm main}$.

The fermion kinetic energy term in (\ref{Lagra}) gives  the 
interaction term 
$- \psi^{\dagger} e(\vec{p}\,' + \vec{p})\cdot \vec{A}/(2m)\psi$ 
in the NRQED Hamiltonian.
Note that, although the NRQED Hamiltonian is not
an expansion into multipoles,
we call this  term the dipole interaction in the following for 
convenience's sake.
Thus  we consider  the Coulomb term  dressed by the transverse photon 
with the dipole couplings. 
The NRQED Feynman rule applied to this diagram gives  
\eqna
&& \psi^{\dagger}(\vec{p}\,')\left({e \over m}\right)^2
i\int^{\Lambda} {d^4k \over (2\pi)^4}
{1 \over (k^0)^2-\vec{k}\,^2  - \lambda^2 + i\epsilon } 
\biggl ( \vec{p}\,'\cdot \vec{p} 
- { \vec{p} \cdot \vec{k}~ \vec{p}\,'\cdot \vec{k} 
\over \vec{k}\,^2 + \lambda^2 }  \biggr )
\nonumber \\
&&{ 1 \over {E+ k^0 -( \vec{p}\,' + \vec{k} )^2 /(2m) + i\epsilon} }
eA^0
{ 1 \over {E+ k^0 -( \vec{p} + \vec{k} )^2 /(2m) + i\epsilon} }
 \psi(\vec{p}).
\endeqna
We chose the contour in the upper half $k^0$ plane to pick up
only the negative energy photon pole.  Then we  neglect $\vec{k}$
in the  kinetic energy term $ ( \vec{p} + \vec{k} )^2 /(2m) $ 
in the electron propagators. This is justified because the
energy transfer between electrons is of order $v^2$ when $|\vec{p}|$
is of order $v$, while the space component of the photon momentum $|\vec{k}|$ 
is of order $v^2$.
%After taking account of the dominant contribution coming from the
%residue of the negative energy pole at  $k^0=-\sqrt{\vec{k}^2+\lambda^2}$
%of the radiative photon, and carrying out the
After this approximation, angular integration over the photon 
momentum $\vec{k}$ becomes trivial, leaving only the $|\vec{k} |$ integration:
\eqna
&& \psi^{\dagger}(\vec{p}\,')\left({e \over m}\right)^2{ 2 \over 3}
\vec{p}\,'\cdot \vec{p}
\int_0^{\Lambda} {dk \over 2\pi^2}
{k^2 \over 2\sqrt{k^2+\lambda^2}}\left(1+{1\over2}
{\lambda^2 \over k^2+\lambda^2}\right)
\nonumber \\
&&{ -2m \over \vec{p}\,'^2 -2 m E + 2 m \sqrt{k^2+\lambda^2}}
 eA^0
{ -2m \over \vec{p}\,^2 - 2 m E + 2 m \sqrt{k^2+\lambda^2}}
 \psi(\vec{p})
\nonumber \\
&&=
\psi^{\dagger}(\vec{p}\,')
 {\alpha \over \pi }{ 8 \over 3}
  \biggl [  \ln\left({2\Lambda \over \lambda}\right)
 -{5 \over 6 } \biggr ]
{ -e \over 8m^2 }(-2\vec{p\,'}\cdot\vec{p})~A^0 ~\psi(\vec{p})
+ {\cal{O}}(v^8).  \label{NRvertex-fermi}
\endeqna
In the last step, we used the on-shell, at-threshold 
condition, $E= -\vec{p}\,^2/ (2m) + {\cal{O}}(v^4)$.

The  diagram with a self-energy on the external electron
line  gives
\eqna
&&{1 \over 2 }\psi^{\dagger}(\vec{p}\,')\biggl [
{-2m \over \vec{p}\,^2 }
\left ({e \over m}\right)^2{ 2 \over 3}
\vec{p}\cdot \vec{p}
\int_0^{\Lambda} {dk \over 2\pi^2}
{k^2 \over 2\sqrt{k^2+\lambda^2}}\left(1+{1\over2}
{\lambda^2 \over k^2+\lambda^2} \right) { -2m \over \vec{p}\,^2- 2mE }
\nonumber \\
&&\{ { -2m \over \vec{p}\,^2 - 2 m E + 2 m \sqrt{k^2+\lambda^2}}
-{ -1 \over  \sqrt{k^2+\lambda^2}} \}
eA^0
+ ( \vec{p} \rightarrow \vec{p}\,')  \biggr ]  \psi(\vec{p})
\nonumber \\
&&= \psi^{\dagger}(\vec{p}\,')
 {\alpha \over \pi }{ 8 \over 3}
\biggl [ \ln\left({2\Lambda \over \lambda}\right)
-{5 \over 6 } \biggr ]
{ -e \over 8m^2 } (\vec{p}\,'^2 +\vec{p}\,^2)~ A^0~
     \psi(\vec{p})
+ {\cal{O}}(v^8).
\label{NRself-fermi}
\endeqna
Note that the term  $ -1 /  \sqrt{k^2+\lambda^2}  $ is the 
$``$mass renormalization term" of NRQED.
\footnote{$``$Tadpole" diagram due to the  
$<\vec{A}\cdot\vec{A}\psi^{\dagger} \psi >$ completely vanishes after
mass renormalization  because this diagram
does not depend on the external fermion momentum.} 

In order to  maintain the equivalence of QED and NRQED
we must include the negative of these contributions  
in $H_{\rm main}$.  (See  Fig. \ref{comparison1}(d).)
>From (\ref{NRvertex-fermi}) and (\ref{NRself-fermi})
we see that this is achieved by adding
the new $``$renormalization" coefficients
to  the Darwin term $-\psi^{\dagger}e\vec{q}\,^2 A^0/(8m^2) \psi$:
\eqn
\psi^{\dagger}(\vec{p}\,')c_{D}^{NRQED}{ -e \vec{q}\,^2 \over 8m^2 }
A^0  \psi(\vec{p})~,
\endeqn
where
\eqn
c_{D}^{NRQED} = {\alpha \over \pi }{ 8 \over 3}
    \biggl [ \ln\left({\lambda \over  2\Lambda }\right)
+{5 \over 6 }\biggr ] ~.
\label{c_qNRQED}
\endeqn
The entire coefficient of the Darwin 
term is the sum of QED and NRQED contributions:
\eqna
c_{D}&=&c_{D}^{QED}+c_{D}^{NRQED}
\nonumber \\
       &=&1 + {\alpha \over \pi}{ 8 \over 3}
    \biggl [ \ln \left({m \over  2\Lambda }\right)
 -{3 \over 8}+{5 \over 6} \biggr ] +2a_e ~.
\endeqna
Actually, this additional contribution from NRQED
serves to eliminate the contribution of the longitudinal polarization
associated with the finite photon mass \cite{french}.
In other words, the $\ln \lambda$ term in
the $``$renormalization" coefficients due to QED is effectively
replaced in the NRQED  $``$renormalization" constant by
\eqn
\ln\lambda \rightarrow \ln(2\Lambda) - {5 \over 6} ~.
\endeqn
Similarly  the NRQED radiative correction to the Fermi term,
$ -\psi^{\dagger}ie \vec{\sigma}\cdot(\vec{q}\times\vec{A})/(2m)\psi$,
yields the correct $``$renormalization"
coefficient of the $W_1$ and $W_2$ derivative Fermi terms, 
$ \psi^{\dagger}ie (\vec{p}\,'^2 +\vec{p}^2) 
\vec{\sigma}\cdot(\vec{q}\times\vec{A})/(8m^3)\psi$
and 
$ \psi^{\dagger}ie (-2 \vec{p}\,'\cdot \vec{p}) 
\vec{\sigma}\cdot(\vec{q}\times\vec{A})/(8m^3)\psi$, respectively,
which 
are  given by
\eqna
c_{W_1}&=& 1+  {\alpha \over \pi}{ 4\over 3}
  \biggl  [ \ln \left({m \over  2\Lambda }\right)
     -{3 \over 8}+{1\over 4} + {5 \over 6} \biggr ]  +  {a_e\over 2} ~,
\nonumber \\
c_{W_2}&=& {\alpha \over \pi}{ 4\over 3}
  \biggl  [ \ln \left({m \over  2\Lambda }\right)
     -{3 \over 8}+{1\over 4} + {5 \over 6} \biggr ]  +  {a_e \over 2} ~.
\endeqna

The radiative correction comes also from  vacuum polarization.
Since  vacuum polarization is a highly virtual process
within the framework of NRQED,
no vacuum polarization term exists  in $H_{\rm main}$.
Instead, its contribution is represented by
the new photon interaction terms in NRQED.
Again we begin with the  nonrelativistic reduction of
QED amplitude with  one vacuum polarization insertion.
QED gives the renormalized vacuum polarization tensor
\eqn
\Pi^{\mu\nu}(q)=(q^{\mu}q^{\nu}-g^{\mu\nu}q^2)\Pi(q^2)~,
\endeqn
with
\eqn
\Pi(q^2)=-{q^2 \over m^2}
\int^1_0 dt{ \rho(t) m^2 \over q^2-4 m^2(1-t^2)^{-1}}~.
\label{pspect}
\endeqn
For the second order, the photon spectral function $\rho_2(t)$
is known to be
\eqn
\rho_2(t)={\alpha \over \pi} { t^2(1-{1\over3}t^2 ) \over 1-t^2 }~.
\endeqn
Expanding $\Pi(q^2)$ around $q^2=0$, we obtain
\eqn
\Pi_2(q^2)=c_{\rm vp}{ -\vec{q}\,^2 \over m^2}  
+ {\cal{O}}(\alpha v^4, \alpha^2 v^2)~,
\endeqn
where
\eqn
c_{\rm vp} = { \alpha \over 15 \pi } ~.
\endeqn
Thus,  in the Coulomb gauge,
two new photon interaction terms are added to the photon  Hamiltonian
\eqn
  c_{\rm vp}A^i(q){ \vec{q}\,^4 \over m^2 }
                   A^j(q)(\delta^{ij}-{q^iq^j \over \vec{q}\,^2}),
\label{vpt}
\endeqn
and
\eqn
  c_{\rm vp}A^0(\vec{q}){ -\vec{q}\,^4 \over m^2 }
                   A^0(\vec{q}).
\label{vpc}
\endeqn

\subsection{Photon-Fermion Scattering Amplitude}
\label{subsec:two-photon}

Let us now turn to the  processes which contain two fermion
operators and two external photons.
To determine the 
$``$ renormalization"  coefficients
we must carry out the nonrelativistic reduction of these
QED scattering amplitudes. 
In practice, however, we don't have to do it 
at all because the $``$renormalization" coefficients of 
these operators up to 
$v^6$ for spin-flip ones and
$v^4$ for spin-non-flip ones
are  identical with  those 
determined  by the scattering amplitude due to
a static external potential
because of  gauge invariance.
For instance, the same $``$renormalization" coefficient $c_S$ for the
spin-orbit interaction term  
\eqn
\psi^{\dagger}
 { i e \over 4m^2} \vec{\sigma}\cdot
            ( \vec{p}\,'\times\vec{p} )A^0 \psi
\endeqn
must be used for both  the seagull term 
\eqn
\psi^{\dagger}
 { -i e^2 \over 4m^2} \vec{\sigma}\cdot
             ( \vec{q_1}\times\vec{A}(q_1)) A^0(q_2) \psi
\endeqn
and  the time derivative term
\eqn
\psi^{\dagger}
 { i e \over 8m^2} q^0\vec{\sigma}\cdot
             (( \vec{p}\,'+\vec{p})\times\vec{A}) \psi~.
\endeqn
The seagull term  is  the only operator involving  two-photons
relevant to our immediate interest.
This contributes to the $(Z\alpha)^2$ and $\alpha(Z\alpha)^2$
corrections.

In general an explicit  nonrelativistic reduction of the
photon-fermion scattering amplitude 
is  necessary only if one wants to find  
the $``$renormalization" coefficients of operators 
of higher order in $v$, such as 
$\psi^{\dagger}\vec{E}\cdot\vec{E}\psi/m^3\sim v^6$.

We note that $H_{\rm main}$  is not an unique
expression. 
Using the equation of motion for the fermion field, we can obtain  
another form of Hamiltonian.
When $H_{\rm main}$ is quantized,
we should be more careful. Use of the equation of motion is
equivalent to the  transformation of the electron field.
We have to take into account the Jacobian of this change
of variables.
Once  the Jacobian is taken into account, two Hamiltonians become 
completely  identical  and  
produce the same results even for the bound state 
calculation \cite{nrqcd}.
This is why we excluded the operators having  time derivatives,
such as $\bar{\psi}(iD_t)^2\psi/m $, from our  consideration, 
since the equation of motion  renders 
$ iD_t\psi$  to $(\vec{p}\,^2/(2m) + {\cal{O}}(v^4))\psi $.  

In this manner  we have obtained  all operators in the main part of
the NRQED Hamiltonian $H_{\rm main}$
necessary for our calculation to the desired order.

\subsection{The NRQED Hamiltonian $H_{\rm main}$}
\label{subsec:Hamil}

For later reference let us write down 
the part of the NRQED Hamiltonian
$H_{\rm main}$ valid to order $\alpha$
by putting together the results of Sec. \ref{subsec:one-photon} 
and \ref{subsec:two-photon}.
To do this, we introduce the $\vec{q}\,^2$ derivative Fermi term 
by combining the $W_1$ and $W_2$ derivative Fermi term at the order $\alpha$.
It is of the form:
\eqna
H_{\rm main}^{\Lambda}  =  \psi^{\dagger}(\vec{p}\,')
        \biggl [ && { {\vec{p}\,^2} \over 2m } + eA^0
           - { (\vec{p}\,^2)^2 \over 8 m^3 }
           -{e \over 2m}  (\vec{p}\,'+\vec{p})\cdot\vec{A}
           +{e^2 \over 2m}  \vec{A}\cdot\vec{A}
\nonumber \\
    &- & {ie \over 2m} c_F\vec{\sigma}
      \cdot (\vec{q}\times\vec{A})
           - {e \over 8m^2} c_D\vec{q}\,^2 A^0
\nonumber \\
   &+&{ i e \over 4m^2} c_S\vec{\sigma}\cdot
                   (\vec{p}\,'\times\vec{p})A^0
    - { i e^2 \over 4m^2} c_S \vec{\sigma}\cdot
             (\vec{q_1}\times \vec{A}(q_1))  A^0(q_2)
\nonumber \\
   &+&{ i e \over 8m^2} c_S q^0 \vec{\sigma}\cdot
                   ((\vec{p}\,'+\vec{p})\times\vec{A})
\nonumber \\
 &+& {ie \over 8m^3} c_W(\vec{p}\,'^2+\vec{p}\,^2)\vec{\sigma}
                   \cdot (\vec{q}\times\vec{A})
     + {ie \over 8m^3} c_{q^2}\vec{q}\,^2\vec{\sigma}
                   \cdot (\vec{q}\times\vec{A})
\nonumber \\
    &+& {ie \over 8m^3} c_{p'p}
               \{ \vec{p}\cdot(\vec{q}\times\vec{A})
               (\vec{\sigma}\cdot\vec{p}\,')
                + \vec{p}\,'\cdot(\vec{q}\times\vec{A})
                     (\vec{\sigma}\cdot\vec{p}) \}
\nonumber \\
    & +&  \ldots~~  \biggr ] \psi(\vec{p})
\nonumber \\
  &+&c_{\rm vp}A^i(q){ \vec{q}\,^4 \over m^2 }
       A^j(q)(\delta^{ij}-{q^iq^j \over \vec{q}\,^2})
\nonumber \\
  &+&c_{\rm vp}A^0(\vec{q}){ -\vec{q}\,^4 \over m^2 }
     A^0(\vec{q})~,
\label{H_main}
\endeqna
where $\vec{p}\,'$ and $\vec{p}$ are  the outgoing and incoming
fermion momenta, respectively, and $q=(q^0,\vec{q})$ is the 
incoming photon momentum. 
In the seagull vertex, $\vec{q_1}$ is
the incoming momentum of the vector potential $\vec{A}$.
The superscript $\Lambda$ indicates that the
Hamiltonian is regularized with the UV cut-off $\Lambda$.
The  $``$renormalization" coefficients are
\eqna
c_F& =&1+ a_e~,
\nonumber \\
c_D&=&1 + { \alpha \over \pi } {8 \over 3}
     \biggl [ \ln \left({m \over 2\Lambda }\right)
-{3 \over 8}+{5 \over 6}\biggr ]
                 + 2 a_e~,
\nonumber \\
c_S&=& 1+ 2a_e ~,
\nonumber \\
c_W&=& 1~,
\nonumber \\
c_{q^2}&=& {\alpha\over\pi} {4 \over 3}\biggl [
\ln\left({m \over 2\Lambda}\right)
- { 3 \over 8 } + { 5 \over 6}  +{1 \over 4}\biggr ]  + {a_e \over 2} ~,
\nonumber \\
c_{p'p}&=& a_e~,
\nonumber \\
c_{\rm vp}&=& {\alpha \over 15\pi } ~.
\label{Rconsts}
\endeqna

Thus far we have not shown explicitly 
the contact term $H_{\rm contact}$ of the NRQED Hamiltonian,
which is also obtained by 
comparison of the electron-muon scattering  amplitudes
in QED and NRQED.
The explicit form of  the contact term will be
given in Sec. III  and IV as we calculate
$\alpha(Z\alpha)$ and $\alpha^2(Z\alpha)$  corrections, respectively,  
to the hyperfine splitting.

\subsection{Application of $H_{\rm main}$ to Bound States}
\label{subsec:bound}

Let us now turn our attention to the bound state calculation
using  $H_{\rm main}$.
The main part of the NRQED Hamiltonian  for the muon field  is
obtained by replacing the charge $e$ by $-Ze$ in the $H_{\rm main}$
for the electron field. For  the nonrecoil hyperfine correction,
only the Fermi and Coulomb terms are necessary in the muon Hamiltonian.
Together with the photon Hamiltonian,  we can construct  various
perturbative potentials appearing in
the nonrelativistic Rayleigh-Schr\"{o}dinger perturbation
theory (\ref{perturb}).
The lowest order contribution $E_F$ to hyperfine splitting comes from
the Fermi potential  (\ref{Fpotential}).
A survey of Eqs. (\ref{H_main}) and (\ref{Rconsts})
shows that the only order $\alpha$ correction
is $a_e E_F$ which exhibits the effect of the $``$renormalization":
$c_F-1= a_e$.
Other possible contributions to the hyperfine splitting coming
from $H_{\rm main}$ are those of  the first order perturbation of
the derivative Fermi term and the seagull term,   and the second order
perturbation which involves   the Fermi term and
the $p^4$ relativistic kinetic term or the Darwin term.
The $(p'^2+p^2)$  derivative Fermi term leads to the potential
of order $(Z\alpha)^6 (m/M) m$: 
\eqn
V_{W}= -{ \pi Z\alpha \over mM }
{(\vec{p}\,'^2+\vec{p}\,^2) \over 4m^2}
(\psi^{\dagger}\vec{q}\times\vec{\sigma}_e\psi )\cdot
(\chi^{\dagger}\vec{q}\times\vec{\sigma}_{\mu}\chi)
{1 \over (\vec{q}\,^2+\lambda^2)  }~.
\endeqn
The Darwin term generates the potential of order $(Z\alpha)^4 m $:
\eqn
V_{D} = {4\pi Z\alpha \over 8m^2}
       (\psi^{\dagger}\psi) (\chi^{\dagger} \chi)
      {\vec{q}\,^2 \over (\vec{q}\,^2+\lambda^2) } ~.
\endeqn

Expectation values of these potentials with respect to the bound state
wave function diverge due to integration
over $\vec{q}$.  This is why we need the help of the contact term
$H_{\rm contact}$ for their cancellation.
When the effect of $H_{\rm contact}$ is included, these four potentials
together give the  $(Z\alpha)^2$ Breit
correction. 
A detailed discussion about the treatment of
these UV divergent operators is found in \cite{thesis} where
the derivation of the Breit $(Z\alpha)^2$ correction from NRQED
is described.

Similarly an $\alpha(Z\alpha)^2$ correction
is obtained when the contribution of the 
$``$renormalization" coefficients is included in each potential.
Third order perturbation theory in $H_{\rm main}$ with an intermediate
radiative photon and  dipole couplings also gives 
the $\alpha(Z\alpha)^2$ correction.
This is because these diagrams have the structure similar to the
derivative Fermi term or the Darwin term as is shown in
the determination of the $``$renormalization" coefficients in NRQED
(See Eqs. (\ref{NRvertex-fermi}) and (\ref{NRself-fermi})).

The additional photon interaction terms (\ref{vpt}) and (\ref{vpc})
due to  vacuum polarization
produce the effective potentials
\eqn
V_{\rm tvp}= { \pi (Z\alpha) \over mM } c_{\rm vp}
{ \vec{q}\,^2 \over m^2 }
(\psi^{\dagger}\vec{q}\times\vec{\sigma}_e\psi )\cdot
(\chi^{\dagger}\vec{q}\times\vec{\sigma}_{\mu}\chi)
{\vec{q}\,^2 \over (\vec{q}\,^2+\lambda^2)^2 }
\label{Vtvp}
\endeqn
and
\eqn
V_{\rm cvp} = -{4\pi Z\alpha  \over m^2} c_{\rm vp}
(\psi^{\dagger}\psi) (\chi^{\dagger} \chi)
{\vec{q}\,^4 \over (\vec{q}\,^2+\lambda^2)^2 } ~.
\label{Vcvp}
\endeqn
We note that the first spin-flip potential $ V_{\rm tvp}$ has exactly
the same structure as the $q^2$ derivative Fermi potential.
Thus it contributes  not  to the order $\alpha(Z\alpha)$
but to the order $\alpha(Z\alpha)^2$.
The spin-non-flip potential $ V_{\rm cvp}$ behaves as a  $\delta$
function potential in the coordinate  space just like
the Darwin potential.
Thus it also contributes to the  $\alpha(Z\alpha)^2$ term through the
second order perturbation theory.

To summarize,  no order $\alpha$ correction exists besides $a_eE_F$, 
where $E_F$ is given by (\ref{E_F}).
In the NRQED formulation, it is transparent
why only the anomalous magnetic moment of a {\it free}  electron
contributes to the order $\alpha$ correction to $ E_F$.

%%%% Sec. 3
\section{The $\alpha (Z\alpha)$ Correction }
\label{sec:KP}

In this section we show how the non-recoil radiative correction
of order $\alpha (Z\alpha)$,
calculated long ago by Kroll and Pollock and by Karplus,
Klein and Schwinger \cite{KPoriginal}, can be obtained within
the framework of NRQED.
For brevity, let us refer to this as the K-P term.
The procedures developed here are readily applicable to
the $\alpha^2 (Z\alpha)$ term calculation in NRQED.

\subsection{Diagram Selection}
\label{subsec:KPdiag}

%%%%
\begin{figure}
\centerline{\epsfbox{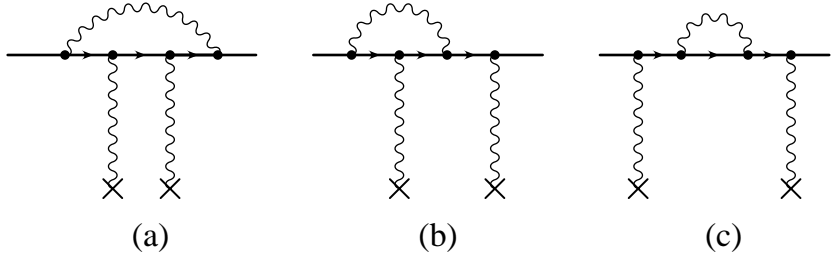}}
\vspace{3ex}
\caption{ QED diagrams contributing to the $\alpha(Z\alpha)$
radiative correction to the muonium hyperfine splitting.
\label{KPqed}}
\vspace{3ex}
\end{figure}
%%%%

The QED diagrams involved in this calculation are shown in
Fig. \ref{KPqed}.
In the original and subsequent works \cite{KPoriginal,BE,yennie,EKS0},
the K-P $\alpha(Z\alpha)E_F$ pure radiative correction was
evaluated from the QED diagrams with the external fermions
put on the mass-shell and at the threshold, and multiplied by
the square of the nonrelativistic  Coulomb wave function
at the origin.
This recipe was justified after complicated and rigorous
consideration of the relativistic bound state theory.
We shall show that NRQED provides an alternative justification
of this procedure in the sense that no other correction term
is needed in this order.

The correction terms whose coefficients are odd powers
of $Z\alpha$ may arise only from very limited sources in the NRQED
bound state theory.
The NRQED Lagrangian $L_{\rm main}$ consists only of terms
with even parity.
This implies that the expectation values of these terms
with respect to the Coulomb wave function are even
in $Z\alpha$, the typical electron momentum
of the Coulomb bound state being $|\vec{p}| \sim (Z\alpha) m $.

The odd power of $Z\alpha$ in the
K-P term $ \sim \alpha(Z\alpha)^5 m^2/M$ therefore
implies that there is no  contribution to it from
the $L_{\rm main}$ part of the NRQED Lagrangian. [Note that the
$``$renormalization" constant $c_i$ in $L_{\rm main}$ does not
depend on $Z\alpha$.
(See Eq. (\ref{Lagra}) and Eq. (\ref{Cdpara})).]

This means that the correction we are looking for must come entirely
from the NRQED contact terms of (\ref{contact}).
To determine the contact term, we compare the scattering amplitudes
evaluated in NRQED and QED in the same power of explicit
$\alpha$ and $Z\alpha$. This comparison is shown in Fig. \ref{comparison2}.
For a given power of the coupling constant $\alpha$,
the number of QED diagrams is
finite while the number of NRQED diagrams is infinite.
We terminate the series of NRQED scattering diagrams using
the power counting rule for their contribution
to  the $bound$ state.

%%%%
\begin{figure}
\centerline{\epsfbox{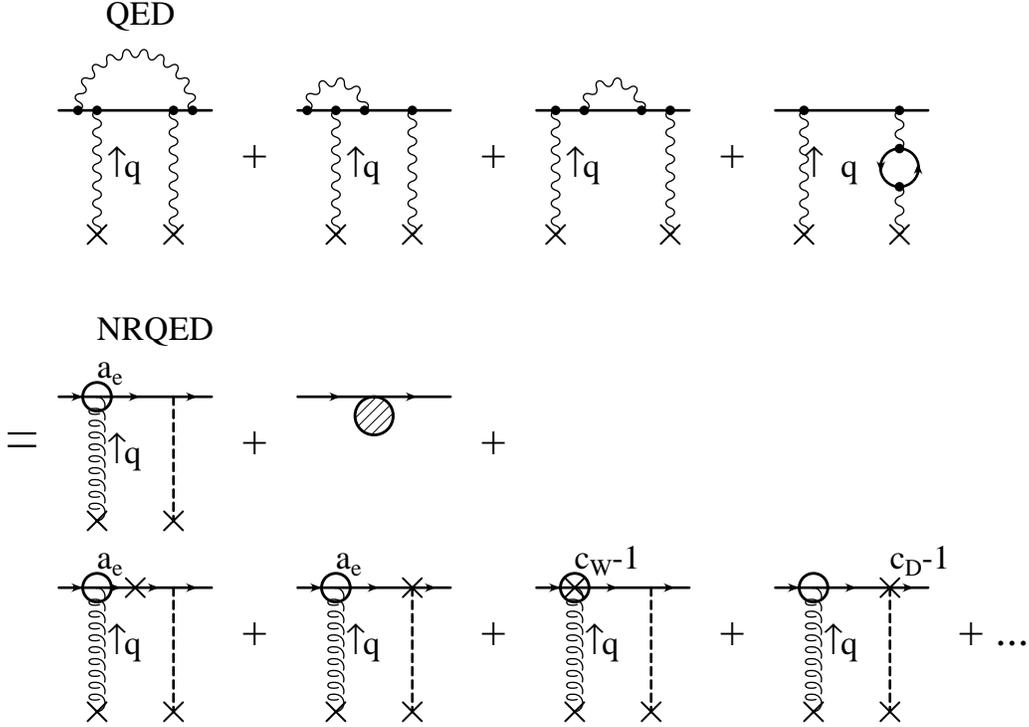}}
\vspace{3ex}
\caption{ QED and NRQED two-photon exchange scattering diagram
comparison in the presence of the radiative correction.
The shaded circle represents the contact term
introduced in this comparison.
The NRQED diagrams in the bottom lines actually contribute to the
$\alpha(Z\alpha)^2$ correction.
\label{comparison2}}
\vspace{3ex}
\end{figure}
%%%%

We chose the electron mass $m$ as the momentum scale of comparison,
and evaluate the scattering amplitudes of both QED and NRQED
on the mass-shell and at the threshold.
In general, this procedure must be carried out for  both
spin-flipping and non-flipping amplitudes. However,
for the K-P term, only the spin-flipping one is needed.
The spin-non-flipping type produces a Lamb-shift type contact term,
which contributes to the hyperfine splitting only
in the order $\alpha(Z\alpha)^7 m^2/M$ and above.

As we have discussed in Sec. \ref{sec:NRQED},
the comparison of QED and NRQED scattering amplitudes
gives rise to a contact term to the NRQED Hamiltonian.
We restrict ourselves to the consideration of
the contact term relevant to the hyperfine
splitting, i.e.,
\eqn
\delta H =  -d_1 {1 \over mM } (\psi^{\dagger} \vec{\sigma_e }\psi)
\cdot(\chi^{\dagger} \vec{\sigma_{\mu} }\chi),
\endeqn
because this is the only source of the K-P
term as was discussed above.

Let us first focus on the contribution
from the vacuum polarization insertion.
The two-photon exchange scattering amplitudes 
containing 
the vacuum polarization potential of (\ref{H_main}) 
contributes to the order
$\alpha(Z\alpha)^2$, not to $\alpha(Z\alpha)$.
Thus the only contribution from the vacuum polarization is
obtained from the contact term which is determined by  calculating
the QED two-photon exchange amplitude
with one vacuum polarization insertion in the photon line
with the on-shell at-threshold  external fermions times
the square of the Coulomb wave function at the origin.

Let us turn next to the contribution from the radiative photon.
The QED diagrams related to this correction are shown in
Fig. \ref{KPqed}.
All three QED scattering amplitudes  have the same form:
\eqn
i{\cal{T}}^{QED} = e^2(Ze^2)^2\int {d^{4}q \over (2\pi)^{4}}
\frac{
\bar{u}_e{\cal E}^{\mu \nu} u_e ~\bar{u}_m{\cal M}_{\mu \nu}
u_m}{(q^{2}+i\epsilon)^{2}}  ~.
\endeqn
Here the electron factor ${\cal E}^{\mu \nu}$ is different for each
diagram but the muon factor ${\cal M}_{\mu \nu}$ is common to
all these diagrams and represents the sum of
the ladder and crossed-ladder diagrams:
\eqn
 {\cal M}_{\mu \nu}=
  \frac{\gamma_{\mu}(\not l - \not q + M)\gamma_{\nu}}
        { (l-q)^{2}-M^{2} + i \epsilon }
+ \frac{\gamma_{\nu}(\not l + \not q + M)\gamma_{\mu}}
        { (l+q)^{2}-M^{2} + i \epsilon }~,
\endeqn
where $l=(M,\vec{0})$ is the external muon momentum and $q$
is the four momentum flowing in the loop between the electron and the muon.
As is well known \cite{SY}, in the limit of infinite muon mass,
the  muon factor reduces to
\eqn
 {\cal M}_{\mu \nu} = {\gamma_{\nu} \not q \gamma_{\mu}}
                    \frac{ -2\pi i \delta(q^{0})}{2M}~.
\endeqn
This is represented by the symbol $\times$ in Figs. \ref{alldiagrams},
\ref{ourdiagrams}, \ref{comparison2}, and \ref{KPqed}.

%
%%%%
\begin{figure}
\centerline{\epsfbox{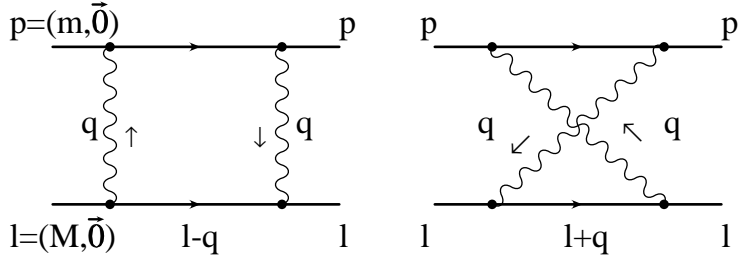}}
\vspace{3ex}
\caption{ Ladder and crossed-ladder diagrams.
\label{muon}}
\vspace{3ex}
\end{figure}
%%%
%

The hyperfine splitting projection operator for strings
of $\gamma$ matrices  is obtained by taking the difference
between the $J=1, J_{z}=0$ state and $J=0$ state and using the
spherical symmetry of the system \cite{CL2}:
\eqn
\frac{1}{12} \sum_{i=1}^{3}
Tr[ {\cal E}^{\mu \nu} \gamma^{5}\gamma^{i}( \gamma^{0}+1)]
  Tr[{\cal M}_{\mu \nu} \gamma^{5}\gamma_{i}( \gamma^{0}+1)]~,
\endeqn
where Roman letters run from one  to three while Greek letters
run from zero to three. This projection is  true  only for
external fermions  on-the-mass-shell and at-threshold.
The trace of the muon factor is easily taken, yielding
\eqn
    \epsilon_{\mu \nu j i} q^{j}
                    \frac{ -2\pi i \delta(q^{0})}{2M}~.
\endeqn
We take the  $\epsilon_{\mu \nu j i}~q^j$ part together with
the electron projection operator as the hfs projection operator, and
the other muon factor will be included as a numerical factor.

In order that these diagrams contribute to the hyperfine splitting,
one of the exchanged photons must be transverse
(attached to a vertex $\gamma^i$)
while the other is Coulombic (attached to a vertex $\gamma^0$).
Our projection operator of hyperfine splitting
picks up automatically this structure from the electron-line.

The corresponding NRQED scattering amplitude consists of many diagrams,
but most of them
actually contribute to the  order higher than the K-P term.
The only  diagram necessary  is a combination of the Fermi potential
multiplied by the NRQED renormalization constant, namely
the second order anomalous magnetic moment,
and the Coulomb potential.
This scattering amplitude is named $i{\cal{T}}^{NRQED}$.

Other diagrams, such as the combination of
the Darwin potential $V_D$ including the $``$renormalization"
constant and the Coulomb potential, have the same power of
explicit $\alpha$ and $Z\alpha$ as the Fermi one,
but  diverges linearly
in both UV and IR region.  These  divergences cancel out in
the bound state calculation.
The detail is similar to the discussion on the Breit term
calculation given in \cite{thesis}.
Eventually they contribute to the terms of order $\alpha(Z\alpha)^2 $.
This argument holds also for the potentials
$V_{\rm tvp}$ and $V_{\rm cvp}$
representing the vacuum polarization effect.

The QED processes with three or more photon exchange
contribute to
obviously higher order terms due to the explicit extra power of
the coupling constant $Z\alpha$.

The contact term can be defined as
the QED amplitude minus the NRQED amplitude 
for the two photon exchange process:
\eqn
 -d_1 {1 \over mM } (\psi^{\dagger} \vec{\sigma_e }\psi)
 \cdot (\chi^{\dagger} \vec{\sigma_{\mu} }\chi )
 \equiv i{\cal{T}}^{QED} - i{\cal{T}}^{NRQED}~. ~~~~\label{difofd_1}
\endeqn
Actually both QED and NRQED amplitudes are IR divergent
in the limit of the vanishing external photon momentum $\vec{q}$.
These threshold singularities cancel each other in the
difference (\ref{difofd_1}).

This contact term is to be put into
the first order perturbation theory. Then the wave function integral
is trivially done, resulting in the square of the Coulomb wave function
at the origin. Thus the K-P term
is given by
\eqn
\Delta \nu({\rm KP})=|\phi(0)|^2 {-d_1 \over mM}
\langle  \vec{\sigma_e } \cdot \vec{\sigma_{\mu}}
\rangle |^{J=1}_{J=0}~~, ~~~\label{KPhfs}
\endeqn
where $|\phi(0)|^2=\gamma^3 / \pi$ for the ground state.
In the actual calculation, we take the difference between spin J=1
and J=0  for the scattering amplitude  first using the projection
operator.

\subsection{Calculation of the QED Amplitude}
\label{subsec:KPcalc}

We have shown that
the $\alpha(Z\alpha)$ non-recoil radiative correction
comes entirely from the NRQED  contact term evaluated at the origin of 
the wave function.  On the other hand, evaluation of the NRQED contact term is 
equivalent to that of the on-shell at-threshold QED scattering amplitude.  
This is why the calculation of the $\alpha (Z\alpha )$ term is 
much simpler than other terms such as the $\alpha (Z\alpha )^2$ term.

Our approach to  carry out the computation of the QED scattering
amplitudes  is by numerical integration.
Let us explain the outline of our procedure.
The detail of the calculation is given in Appendix \ref{KPnum}.
The electron-line structure of  each diagram
is directly written down  using the
parametric Feynman-Dyson rules for QED \cite{CvK,kinoshitabook}.
Feynman parameters assigned to the
electron-line are $z_1$, $z_2$, and   $z_3$, while one assigned
to the radiative photon line is $z_4$.
The momenta flowing in the fermion lines after the
radiative photon loop momentum is integrated out are expressed in terms of
correlation functions  $B_{ij}$, which are functions of Feynman parameters
and  determined by the topology  of the loop
structure of the diagram alone. Then our integrals are expressed as two or
three dimensional Feynman-parametric integrals with an additional
one dimension corresponding to the magnitude of the momentum
$\vec{q}$ of the external potential.

Two of the QED diagrams have UV divergences and must be renormalized.
The renormalization terms are  generated  using the
projection operators in the algebraic program FORM \cite{form}.
Our projection operators for QED renormalization
constants are quite general and  applicable to any order.
They are presented in Appendix \ref{projection}.
All of the renormalization constants are determined in the on-shell
scheme.
These renormalization terms should be expressed by the same
Feynman parameters as those assigned to the original diagrams
in order to realize point-by-point subtraction in numerical
integration by means of the adaptive
iterative Monte-Carlo integration routine VEGAS \cite{lepage}.

The hfs contribution due to the second order anomalous
magnetic moment should be subtracted from the diagrams
involving the second order vertex correction.
Actually it is very easily done along with the charge
renormalization:
Let the external photon momentum $\vec{q}$ tend to zero in the original
diagram expression of the electron factor.
Then subtract this IR limit from the original diagram.
We can easily prove that
this IR limit of the diagram is nothing but the
sum of the  charge renormalization constant and
the anomalous magnetic moment of the second order.
(See \cite{thesis} for details.)

Even though all diagrams are free from UV divergences after
the renormalization is completed, they still suffer from IR divergence.
In general, the Coulomb bound state
has two kinds of IR divergence: one is due to
the threshold singularity, and the other is due to
the radiative photon.

The mechanism of threshold singularity is the following.
In order to contribute to the hyperfine splitting,
one of the two exchanged photons must be Coulomb-like while
the other is transverse.
This Coulomb photon may be absorbed in the wave function.
As a result, the diagram is reduced to one of lower order
in $Z\alpha$, or multiplied by $1/(Z\alpha)$. This is  the physical
origin of this type of IR divergence, which is  ubiquitous
in the relativistic treatment of bound state problem.
In the calculation of the $\alpha(Z\alpha)$ correction,
however, such a $``$divergence" can be avoided completely by
subtracting  the contribution of the
free anomalous magnetic moment.
This is because other threshold singularities
are absent due to the on-shell renormalization.

The remaining IR singularity is caused by radiative photons.
Our choice to deal with this singularity is to put a
small photon mass $\lambda$ in the radiative photon.
For the K-P term, this singularity must cancel out
when all QED diagrams of the  gauge invariant set are included.

\subsection{Summary of the $\alpha(Z\alpha)$ Correction}
\label{subsec:KPsumm}

We have shown that non-recoil radiative corrections to the muonium
hyperfine splitting having the odd power of $Z\alpha$
comes only from the contact term of NRQED,
and that this contact term is determined  as
the difference between the QED and NRQED scattering amplitudes.

The resulting expression for the K-P radiative correction
can be evaluated either analytically or numerically.
We have chosen the later approach.
The three dimensional integration has been carried out
using the adaptive
iterative Monte-Carlo integration routine VEGAS~\cite{lepage}.
Each diagram has the IR divergence of the form
proportional to $\sqrt{m / \lambda}$,  but their sum is finite.
Our numerical evaluation shows
that the contribution due to the radiative photon  is given by
\eqn
 \Delta \nu({\rm KP})_{\rm ph}  = -2.556~80(6) \alpha(Z\alpha) E_F.
\endeqn
We have also evaluated this integral analytically and obtained the
same result as that of
Kroll and Pollock,
and Karplus, Klein, and Schwinger \cite{KPoriginal}:
\eqn
 \Delta \nu({\rm KP})_{\rm ph}
      =  \biggl (\ln 2 - {13 \over 4 }\biggr) \alpha(Z\alpha) E_F
=  (-2.556~852 \cdots )\alpha(Z\alpha)E_F~.
\label{KP_photon}
\endeqn
An  easy analytic calculation of the vacuum-polarization contribution gives
\eqn
\Delta \nu({\rm KP})_{\rm VP}= {3 \over 4}\alpha(Z\alpha) E_F.
\label{KP_vp}
\endeqn
Putting these results together we obtain
the well known $\alpha(Z\alpha)$ correction
in the frame work of NRQED:
\eqn
\Delta \nu({\rm KP})
      = \biggl (\ln 2 - {13 \over 4 }+{3 \over 4} \biggr)
    \alpha(Z\alpha) E_F~.
\endeqn
This justifies the procedure adopted in Ref.\cite{KPoriginal}.

%%%% Sec.4
\section{The $\alpha^2(Z\alpha)$ correction}
\label{sec:KN}

\subsection{Diagram Selection}
\label{subsec:KNdiag}

In this section,
we  give an outline of the evaluation of
the $\alpha^{2}(Z\alpha) $ correction
to the Fermi frequency $E_F$ which comes from the six gauge
invariant sets
of  QED Feynman diagrams represented by Fig. \ref{alldiagrams}.
Our treatment of the bound state
to find the contribution to hyperfine splitting
coming from these diagrams is completely identical with that
of the $\alpha(Z\alpha)$ K-P correction.
A new diagram appearing in this order is the light-by-light
scattering insertion.
The light-by-light scattering is  a high energy process
in NRQED. Thus it is represented  only by a contact term in NRQED.
As a result we have to include  the four-photon interaction in
the NRQED Hamiltonian. But as an operator it contributes to
orders higher than our interest here.
Therefore,  what to do is again to calculate the contact term
starting from the scattering
amplitudes of these diagrams with the on-shell at-threshold
particles and
then subtract the contribution of the fourth order anomalous
magnetic moment from Figs. \ref{alldiagrams}(d) and \ref{alldiagrams}(f).

The numerical evaluation of Fig. \ref{alldiagrams} (a) -- (e)
can be carried out easily and  our results are consistent with
those previously obtained  by Eides and his
collaborators \cite{EKS2,EKS3,Eides}.
In contrast, the diagrams of Fig. \ref{alldiagrams} (f)
require a substantial  effort to compute.
A complete evaluation of this contribution is  the main result
of this paper.

\subsection{Calculation of the QED Amplitude}
\label{subsec:KNcalc}

Let us now discuss some technical details of
calculation of (\ref{newresult}) represented by the nineteen diagrams of
Fig. \ref{ourdiagrams}.
Since the bound state structure of these diagrams is identical with
that of the $\alpha(Z\alpha)$ correction,
the procedure of numerical evaluation
of the $\alpha(Z\alpha)$ correction
given in Appendix \ref{KPnum} 
can be applied readily to these  diagrams.
We applied numerous techniques developed for the numerical calculation
of the  anomalous magnetic moment $g-2$ of
the electron \cite{kinoshitabook}, except that we avoided the use of
$``$intermediate" renormalization which was introduced in the $g-2$
calculation to avoid the IR singularity of each diagram.
Instead we use the conventional renormalization procedure which is
IR singular
in the radiative photon mass $\lambda$. This is because
these IR divergent terms are needed to cancel out the other IR singularity,
the threshold singularity in the vanishing external photon
momentum $\vec{q}=0$,  in the proper diagram.
The detail of this mechanism is described in the calculation
of the $\alpha(Z\alpha)$ K-P correction in Appendix \ref{KPnum}.

It is convenient to divide
the nineteen diagrams into four groups:
\newline
Group 1: Diagrams containing  fourth-order vertex corrections.
They are represented by the diagrams
$H_{01}, H_{02}, H_{03}, H_{09}, H_{10}$ and $H_{11}$ of Fig.
\ref{ourdiagrams}.
\newline
Group 2: Diagrams containing fourth-order self-energy insertions.
They are represented by the diagrams
$H_{04}, H_{12}$ of Fig. \ref{ourdiagrams}.
\newline
Group 3: Diagrams in which radiative photons span over two
external photons.
They are represented by the diagrams
$H_{05}, H_{06}, H_{07}, H_{08}, H_{13}, H_{14}, H_{15}$, and $H_{16}$ of
 Fig. \ref{ourdiagrams}.
\newline
Group 4: Diagrams containing
two non-overlapping second-order radiative corrections.
They are represented by the diagrams
$H_{17}, H_{18},$ and $H_{19}$  of Fig. \ref{ourdiagrams}.

The integrands corresponding to the individual diagrams of
Fig. \ref{ourdiagrams} were initially generated using
the algebraic program SCHOONSCHIP \cite{schoon}.
Later we generated the same integrands by FORM \cite{form} as a check.

The parametric representation of Group 1 diagrams is of the form
\eqn
 {3 \over 32}
{\alpha^2(Z\alpha) \over \pi}E_F{m \over \pi^2}
\int^{\infty}_0 {dq \over \vec{q}\,^2}
   {\bf F_1^{\mu}}
 \int{(dz)_{1-4} \over U^2}{1 \over V^2}
 \biggl [{\not p + \not q + 1 \over -\vec {q}^2 }\biggr ]
   \gamma^{\nu}~,
\label{G1}
\endeqn
where the diagram $H_{01}$, for example, has the electron-line operator
\eqn
    {\bf F_1^{\mu}}=\gamma^{\alpha}(\not D_1+m)\gamma^{\beta}(\not D_2+m)
             \gamma_{\beta}
              (\not D_3 +m)\gamma^{\mu}(\not D_4+m)\gamma_{\alpha}~.
\endeqn
Other diagrams of this group are obtained by  permutation of $\gamma$ matrices.
(See Ref. \cite{kinoshitabook} for the definition of $U$, $V$,
$D_i$, etc.)
Using the hyperfine splitting  projection operator, 
one finds that the terms contributing to the hyperfine splitting
are proportional to at least $\vec{q}\,^2 $,
and  kills one of the $\vec{q}\,^2$'s  in the denominator in Eq. (\ref{G1}).
Thus Eq.(\ref{G1}) leads to the energy shift of the form
\eqn
\Delta \nu_{G1} =
{\alpha^2(Z\alpha) \over \pi}E_F { m \over \pi^2}
  \int^{\infty}_0 {dq \over (-\vec{q}\,^2)}
 \int{(dz)_{1-4} \over U^2}\biggl [
{F_0 \over V^2} + {F_1 \over UV} + {F_2 \over U^2V^0} \biggr ]~ ,
\endeqn
where $1/V^0$ is a symbolical representation of  $-\ln V$
in which  the UV divergence is
regularized and subtracted by the corresponding counterterm.
(See Appendix \ref{KPnum} for  a precise definition.)

The parametric representation of Group 2 diagrams is of the form
\eqn
- {3 \over 32}
{\alpha^2(Z\alpha) \over \pi}E_F { m \over \pi^2}
 \int^{\infty}_0 {dq \over \vec{q}\,^2}
   \gamma^{\mu}
   \biggl [{\not p + \not q + m \over -\vec{q}\,^2 }\biggr ]
  {\bf F_2}
 \int{(dz)_{2-4} \over U^2}{1 \over V}
   \biggl [{\not p + \not q + m \over -\vec{q}\,^2 }\biggr ]
   \gamma^{\nu}~,
\endeqn
where, for example, the diagram $H_{04}$ has the electron-line operator
\eqn
    {\bf F_2}=\gamma^{\alpha}(\not D_2+m)\gamma^{\beta}(\not D_3+m)
                \gamma_{\beta}(\not D_4+m)\gamma_{\alpha}~.
\endeqn
Its contribution to the hyperfine splitting has the form
\eqn
\Delta \nu_{G2}=
{\alpha^2(Z\alpha) \over \pi}E_F { m \over \pi^2}
\int^{\infty}_0 {dq \over (-\vec{q}\,^2)^2}
 \int{(dz)_{2-4} \over U^2}\biggl [{F_0 \over V} 
+{F_1 \over UV^0} \biggr ] ~.
\endeqn

For the diagrams $H_{04}$ and $H_{12}$, the product of
two electron propagators
just outside the fourth order self-energy diagram behave as
$(1/\vec{q}\,^2)^2$, which makes the convergence of the numerical
integrals difficult
in the small $|\vec q|$ region, even though the integrals are
analytically free from
the IR singularity in $|\vec{q}|$
after the mass and wave function renormalizations are carried out.
In order to avoid this computational difficulty,
we introduced an additional parameter $y$ varying from zero to one to 
combine the original term and the renormalization term.
All the numerator expressions are then proportional to at least
$\vec{q}\,^{2}$ and  kills one of the electron propagators.

The parametric representation of Group 3 is of the form
\eqn
- {3 \over 16}
{\alpha^2(Z\alpha) \over \pi}
E_F{ m \over \pi^2}
\int^{\infty}_0 {dq \over \vec{q}\,^2}
     {\bf F_3^{\mu,\nu}} \int{(dz)_{1-5} \over U^2}{1 \over V^3}~.
\endeqn
For instance, the diagram $H_{05}$ has  the electron-line operator
$\bf F_3^{\mu,\nu}$:
\eqn
 {\bf F_3^{\mu,\nu}}=\gamma^{\alpha}(\not D_1+m)
                    \gamma^{\beta}(\not D_2+m)
                   \gamma_{\beta}(\not D_3+m)\gamma^{\mu}
    (\not D_4+m)\gamma^{\nu}(\not D_5+m)\gamma_{\alpha}~.
\endeqn
By using the hyperfine splitting  projection operator, we get
\eqn
\Delta \nu_{G3} =
{\alpha^2(Z\alpha) \over \pi} E_F{ m \over \pi^2} \int^{\infty}_0 dq
      \int{(dz)_{1-5} \over U^2}
 \biggl [{F_0 \over V^3}+ {F_1 \over UV^2} +{F_2 \over U^2V} \biggr ]~.
\endeqn

The Group 4 diagrams $H_{17}, H_{18},$ and $H_{19}$ contain
two non-overlapping second-order radiative corrections.
Their sum is invariant under the covariant gauge transformation
and free from the IR singularity due to the radiative photons.

Let us consider the sum of two diagrams of Fig. \ref{ir1789}.
If the photon propagator is chosen as
\eqn
     { -i \over k^2 + i\epsilon }
           ( g^{\mu\nu}-\beta { k^{\mu}k^{\nu} \over k^2 } )~,
\label{gauge}
\endeqn
the gauge-dependent part of the vertex diagram gives
\eqna
\beta \int^\Lambda {d^4k \over (2\pi)^4 } & \not k&
{ i \over \not p - \not k + \not q - m}
          \gamma^{\mu} { i \over \not p - \not k  - m}
          \not k { -i \over k^2 }
\nonumber \\
& = & i \beta \int^{\Lambda}{d^4k \over (2\pi)^4 k^2 }\not k
 { 1 \over \not p - \not k + \not q - m}
          \gamma^{\mu} (- 1)~,
\label{lgtvex}
\endeqna
where we use the on-shell condition $\not p=m$.
The self-energy diagram with one external photon diagram is
\eqna
\beta \int^\Lambda {d^4k \over (2\pi)^4 } &\not k &
{ i \over \not p - \not k + \not q - m}
          \not k { i \over \not p + \not q  - m}
          \gamma^{\mu} { -i \over k^2 }
\nonumber \\
& = & i \beta \int^{\Lambda}{d^4k \over (2\pi)^4 k^2 }\not k
 \biggl [{ 1 \over \not p - \not k + \not q - m} \gamma^{\mu}
 -{ 1 \over \not p + \not q - m} \gamma^{\mu} \biggr ]~.
\label{lgtsen}
\endeqna
The second term, which is related to the mass renormalization
constant proportional to  the longitudinal
photon  polarization, vanishes  when the integration
over $k$  is carried out with a proper regularization.
Then the gauge dependent parts of
(\ref{lgtvex}) and (\ref{lgtsen}) cancel each other and the sum
is independent of particular choice of gauge.

%
%%%
\begin{figure}
\centerline{\epsfbox{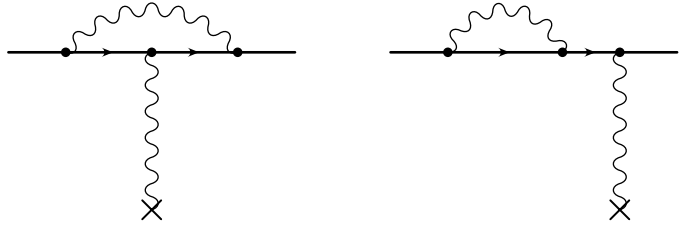}}
\vspace{3ex}
\caption{Sum of the second-order self-energy diagram
and vertex diagram.
\label{ir1789}}
\vspace{3ex}
\end{figure}
%%%

The numerical integration is performed for the integral combining
three diagrams $H_{17}, H_{18}$ and $H_{19}$
together so that  cancellation of IR divergences
occurs in the same region of the  Feynman parametric space.
The result  obtained for the zero mass radiative photon
($\lambda^2=0$)  is
\eqn
\Delta \nu{(H_{17})}+\Delta\nu{(H_{18})}+\Delta\nu{(H_{19})} =
 -0.478~03~(15)  {\alpha^2(Z\alpha)\over \pi} E_F  ~.
\endeqn
This is in good agreement  with  the result calculated
in the Fried-Yennie gauge \cite{EKS4},
in which $\beta=-2$ in (\ref{gauge}),
\eqn
\Delta \nu{(H_{17})}+\Delta\nu{(H_{18})}+\Delta\nu{(H_{19})} =
 -0.477~89~(1)  {\alpha^2(Z\alpha)\over \pi} E_F ~.
\endeqn
Note that $\Delta \nu{(H_{17})}$, $\Delta\nu{(H_{18})}$,
and $\Delta\nu{(H_{19})}$ individually are gauge dependent, and their values
are completely different between our results and those of Ref. \cite{EKS4}.

\subsection{Problems Concerning Numerical Integration}
\label{subsec:KNint}

\begin{table} [p]
\centerline{\epsfbox{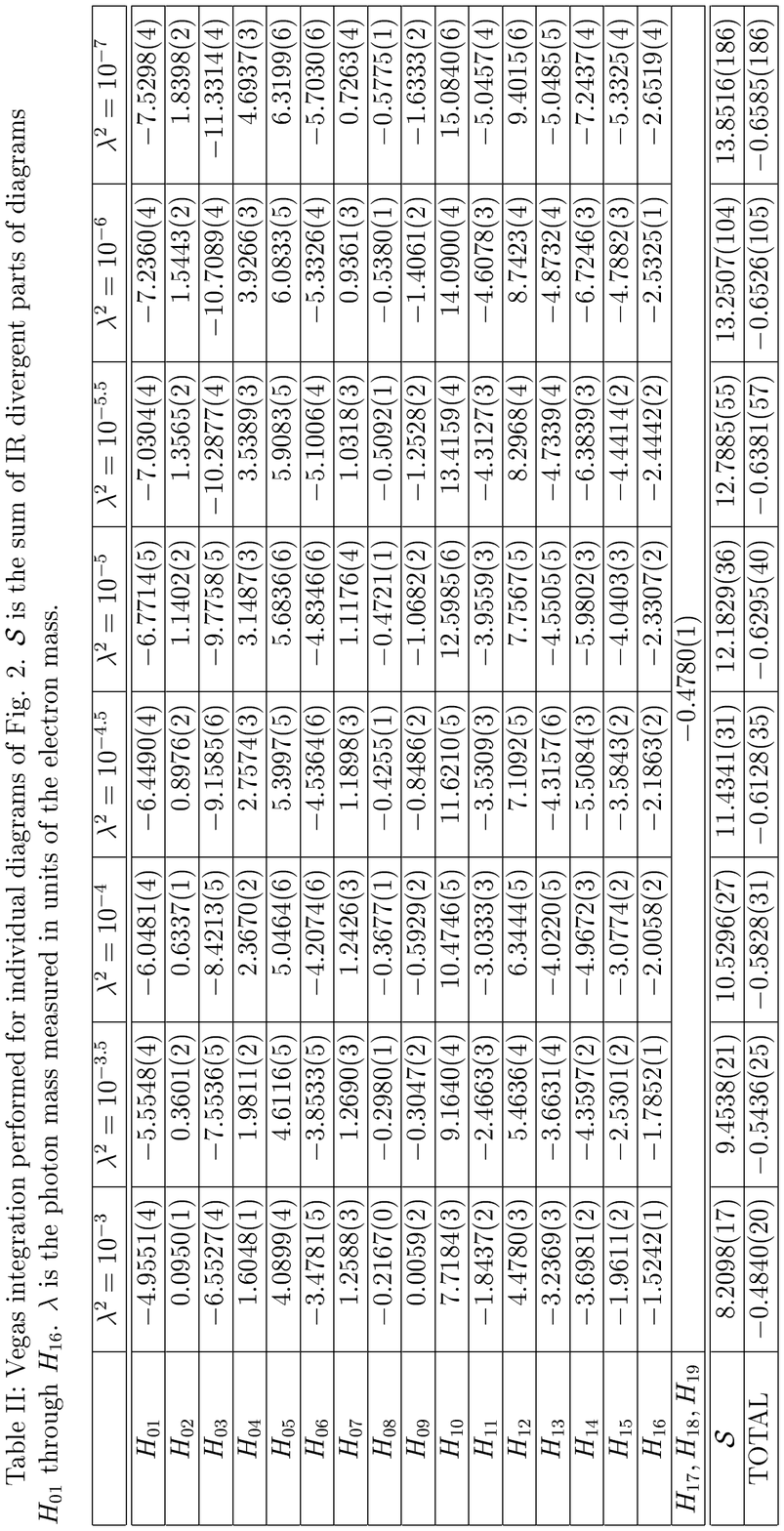}}
%\caption{ VEGAS integration performed for individual diagrams of Fig.2.
%${\cal S }$ is the sum of IR divergent parts of diagrams
%$H_{01}$ through $H_{16}$.
%$\lambda$ is the photon mass measured in units of the electron mass.
%[Should we expand this Table to include results for larger $\lambda$ ?]
%\label{table1}}
%\vspace{3ex}
%\input{table1.tex}
\end{table}

Let us now discuss some technical details of
calculation of $\Delta \nu(H_{01})$ to $\Delta \nu(H_{16})$.
After the ultraviolet divergences are renormalized,
individual diagrams still suffer from severe infrared (IR) divergence,
which is of the form $\lambda^{-1/2}$, $\lambda$ being the photon rest 
mass measured in units of the electron mass.
Of course, the sum over all diagrams of Fig. \ref{ourdiagrams}
is free from the IR divergence.
This does not mean, however, that the sum can be integrated easily on
a computer.
This is because the IR finiteness results from cancellation of divergences
for $\lambda \rightarrow 0$ from different parts of the integration domain.

One way to deal with this problem is to evaluate individual integrals
for several small values of $\lambda$ and extrapolate the sum of all
terms to zero photon mass.
Unfortunately, this approach creates integrals of
order $10^3$ for $\lambda^2 \sim 10^{-7}$, while their sum is of order 1,
making it very difficult to control the numerical accuracy of the result.
Another way is to integrate, for $\lambda \neq 0$, the sum of all terms,
which enables us to avoid dealing directly with large numbers.
This approach will also result in a better error estimate.
The main practical difficulty is the large amount of computing time required.

\begin{figure}
\caption{
The graph  for the coefficient $y$ of
$E_F \alpha^2(Z\alpha)/ \pi$ obtained by VEGAS  versus
$x\equiv\lambda^{1\over2}$.
The solid line is  the chi-square fit for $ y=a_0 + a_1 x + a_2 x^2 $, where
$a_0, a_1$, and $a_2$  are determined from the data  listed in Table I. }
\centerline{\epsfbox{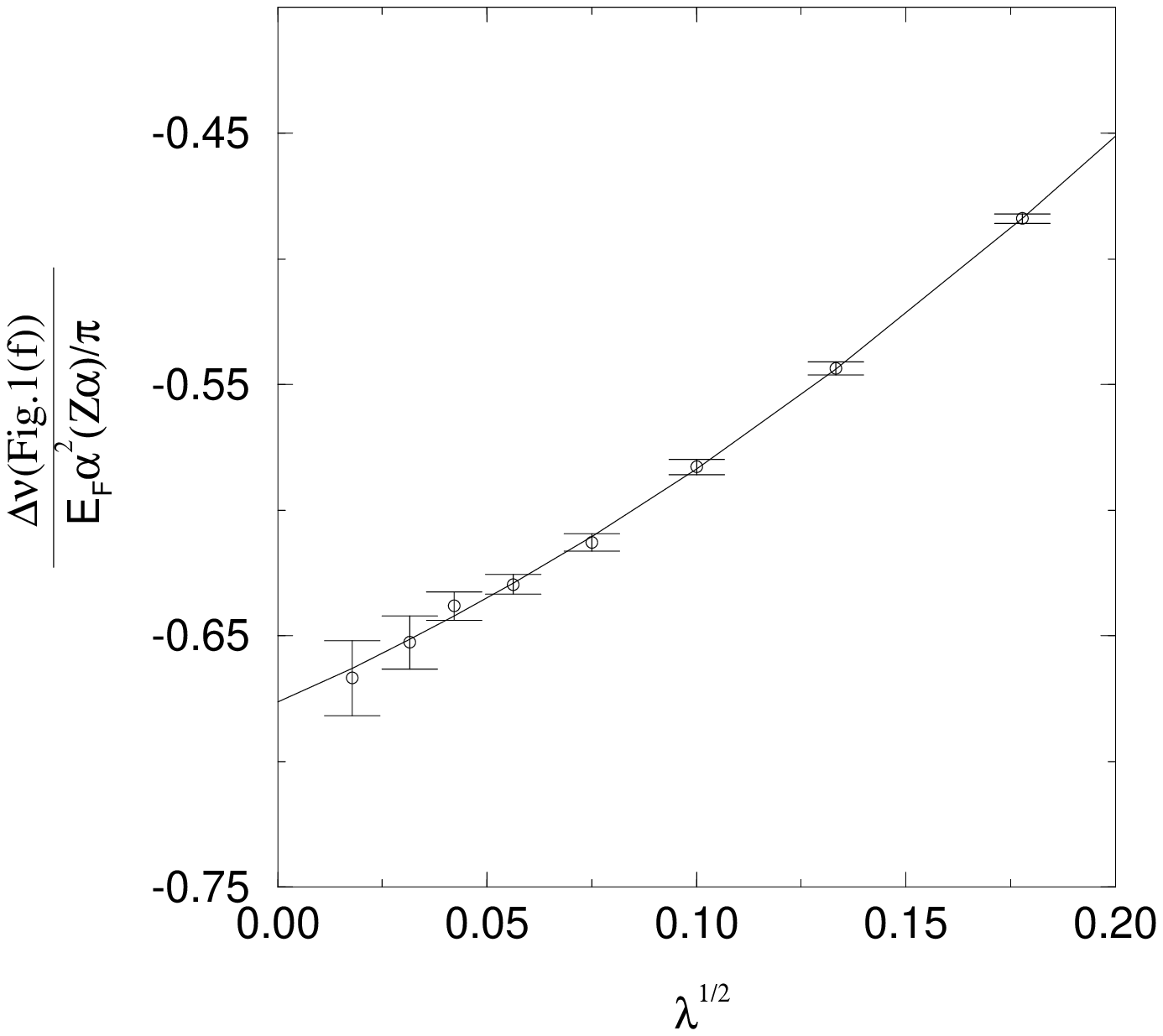}}
\label{kngraph}
\end{figure}

This problem can be somewhat alleviated if one evaluates each integral
after subtracting its IR-divergent part, and then evaluates
the sum $\cal S$ of the IR-subtraction terms of all diagrams.
This method, which we have chosen, ensures that all integrals stay small
(less than $\sim$ 20) for any value of $\lambda$.
Thus far, we have evaluated them for several values of $\lambda^2$ in
the range of $10^{-3}$ to $10^{-7}$.
The integration has been carried out numerically using the adaptive
iterative Monte-Carlo integration routine VEGAS~\cite{lepage}.
The result of each integration is summarized in  Table II.
The degree of difficulty of numerical integration for $\cal S$
increases rapidly as $\lambda$ decreases.
This prevents us from going to smaller values of $\lambda$ at present.
Even $\lambda^2 = 10^{-7}$ is a struggle.
Although evaluation of integrals up to $\lambda^2 = 10 ^{-8}$ is highly
desirable, we have not attempted it thus far since
it will require an extraordinary amount of computing time.

%%%%%
\addtocounter{table}{1}
\begin{table}
\caption{ The chi-square fitting for the coefficient $y$ of
$E_F \alpha^2(Z\alpha)/ \pi$ versus $x\equiv\lambda^{1\over2}$
where $y = a_0 + a_1 x^2 + \cdots$ and $\lambda$ is the photon mass
measured in units of the electron mass.
\label{fit} }
\[
\begin{array}{ccccc}
\hline
 \multicolumn{1}{c}{{\rm photon~mass}~\lambda^2}
&\multicolumn{1}{c}{}
&\multicolumn{1}{c}{ y=a_0+a_1 x }
&\multicolumn{1}{c}{y=a_0+a_1 x+a_2 x^2 }
&\multicolumn{1}{c}{y=a_0+a_1 x+a_2 x^2 +a_3 x^3}
\\ \hline
              &  a_0  & -0.678\pm 0.015 &  -0.700\pm0.041  & \mbox{-----}
\\
10^{-7} \leq \lambda^2 \leq 10^{-5}  &  a_1  &   0.88\pm0.29
                                        & 2.03\pm2.05  & \mbox{-----}
\\
              &  a_2  &\mbox{-----}  &  -13.77\pm24.06  & \mbox{-----}
\\ \hline
              &  a_0  & \mbox{-----} &  -0.6764\pm0.0079  & -0.672\pm0.017
\\
10^{-7} \leq \lambda^2 \leq 10^{-3}  &  a_1  & \mbox{-----} &  0.73\pm0.15
                                                              & 0.58\pm0.57
\\
              &  a_2   & \mbox{-----} &     1.99\pm0.63   &  3.45\pm5.62
\\
              &~  a_3 ~  & \mbox{-----} &  \mbox{-----}   &  -4.39\pm 16.78
\\ \hline
\end{array}
\]
\label{table2}
\end{table}
%%%%%%

The data in Table II shows that the contribution of Fig. 1(f)
falls within errors on a straight line for 
$10^{-7} \leq \lambda^2 \leq 10^{-5}$.
Thus the extrapolation to $\lambda=0$ may be tried with
a linear polynomial $a_0 + a_1 x$, where $x=\lambda^{1/2}$ \cite{numrecipes}.
The upper box of column 3 of Table \ref{table2} 
shows the best linear fit to the data in this range of $\lambda$.

We also list in the upper box of column 4 a fit to the same set of 
data in terms of a quadratic polynomial 
$a_0+a_1x+a_2x^2$.
Clearly the result is much less certain than the linear fit.
This is because the increased flexibility of the quadratic polynomial is 
now responding not only to the nonlinearity of data but also
to the noise of numerical integration.
For this reason the fitting with a linear polynomial will be more 
appropriate for this data.

On the other hand, if one tries to fit the entire data of Table II
which show clear deviation from linearity, the linear fit is no 
longer appropriate 
and one must use (at least) a 
quadratic polynomial (See Fig. 9).  The best fit by a quadratic 
polynomial to the whole set 
of data of Table II is shown in the  bottom box of column 4.  
The bottom box of column 5 shows that 
use of a cubic polynomial is not recommended to this data because it 
responds more to the noise of numerical integration than to the real signal.
Based on these considerations we believe that 
the value of $a_0$ determined by the quadratic fit gives the best estimate
of $ \Delta \nu(\mbox{Fig.1(f)})$: 
\eqn
 \Delta \nu(\mbox{Fig.1(f)}) = -0.676~4~(79) 
      {\alpha^2(Z\alpha) \over \pi }E_F~ .
\label{KN2result}
\endeqn
%
%

%%%% Sec. 5
\section {Discussion}
\label{sec:DISCUSS}

The remaining uncertainty in the value of $\Delta \nu(\mbox{Fig.1(f)})$
is still considerable.
Nevertheless it is a factor 5 improvement over the
preliminary value (\ref{KN1result}).
Since we published the preliminary result,
Eides $et.~al$. \cite{EKS4,ES1} have completed their calculation and
reported a more accurate value
\eqna
  \Delta \nu (\mbox{Fig.1(f)}) &=& -0.671~1~(7) {{\alpha^2 (Z\alpha )}
\over \pi} E_F
 \nonumber  \\
&=& -0.370~1~(4)~\mbox{kHz} . ~~~~~~~~~~~~~~~~~~~~\label{Eidesnewresult}
\endeqna
Our new result (\ref{KN2result}) is in close agreement with
the result of Ref.\cite{ES1}.
Note that the calculation of Ref. \cite{ES1} is carried
out in the Fried-Yennie gauge while our work is carried out in the
Feynman gauge.
Considerably higher precision of (\ref{Eidesnewresult})
over (\ref{KN2result}) 
reflects the advantage of calculation in the Fried-Yennie gauge 
which renders each diagram IR-divergence-free 
when it is transformed by application of integration by part.
This approach requires nontrivial amount of diagram-specific manipulation
because of very complicated integrands involving many variables.
In the Feynman gauge calculation, on the other hand,
each diagram is IR divergent due to the retarded Coulomb-like photons,
which makes individual integral cutoff dependent.
The advantage of this approach is that one can apply a systematic computer 
algebraic method, which minimizes the chance of making mistakes $-$ an 
important consideration in such a complicated calculation.
The agreement of (\ref{KN2result}) and (\ref{Eidesnewresult}) confirms 
gauge independence of the result
to the extent of numerical precision.

The total contribution of order $\alpha^2(Z\alpha)$ including the results
(\ref{fig1a})--(\ref{ourfig1e}) and (\ref{KN2result})
is given in (\ref{alpha^2(Zalpha)}).
%\eqna
%\Delta \nu (\mbox{Fig.1})  &=& ~0.769~7~(81) {{\alpha^2 (Z\alpha )}
%\over \pi} E_F
% \nonumber  \\
%&=& ~0.424~(4)~\mbox{kHz} . ~~~~~~~~~~~~~~~~~~~~\label{alpha^2(Zalpha)}
%\endeqna
If we instead use (\ref{Eidesnewresult}) for Fig. \ref{alldiagrams}(f),
the total contribution becomes
\eqna
\Delta \nu (\mbox{Fig.1})  &=& ~0.773~2~(7) {{\alpha^2 (Z\alpha )}
\over \pi} E_F
 \nonumber  \\
&=& ~0.426~4~(4)~\mbox{kHz} . ~~~~~~~~~~~~~~~~~~~~\label{Eidesa^2(Za)}
\endeqna
If we add the $\alpha(Z\alpha)^2 $ correction to the previous theoretical
prediction (\ref{oldtheory}), we obtain
\eqn
\Delta \nu (\mbox {new theory} )= 
\left \{ \begin{array}{ll}
4~463~302.69~(1.34)~(0.21)~(0.16)~\mbox {kHz}
                        &~~~~~~\mbox{Eq.} (\ref{alpha^2(Zalpha)}) , 
\\
4~463~302.70~(1.34)~(0.21)~(0.16)~\mbox {kHz} 
                        &~~~~~~\mbox{Eq.} (\ref{Eidesa^2(Za)}) .
          \end{array}
                 \right .
\label{newtheory}
\endeqn
Now that the complete $\alpha^2(Z\alpha)$ correction has been evaluated, 
the major remaining source of theoretical uncertainty in
the muonium hyperfine structure is, as is seen from (\ref{nonrecoil}),
the numerically evaluated nonlogarithmic part of the 
$\alpha(Z\alpha)^2$ correction term.
In addition, 
although not listed in (\ref{newtheory}),
the leading logarithmic corrections of
order $\alpha^{4}$ and  $\alpha^{3}(m/M)$ 
turn out to contribute to the hyperfine structure
as much as the $\alpha^2 (Z\alpha)$ term,
as was shown in our preliminary report \cite{KN1} and also by Karshenboim
\cite{Karsh}. 
The parts of these higher order terms evaluated thus far add up to $-0.68~(6)~$ kHz
if some errors in \cite{KN1} and
\cite{thesis} are corrected.
(Details will be discussed in subsequent papers.)
We cannot simply add these corrections to (\ref{newtheory}), however,  
because the previous evaluation of $\alpha(Z\alpha)^2$ term \cite{sapirstein}
contains parts which are of higher order in $Z\alpha$. 
We have evaluated the $\alpha (Z\alpha )^2$ and $\alpha (Z\alpha )^3$ terms separately in the NRQED formalism in order to 
avoid possible double counting and
reduce the uncertainties due to these terms.
The results will be reported in the subsequent papers.

\acknowledgments

We thank G. P. Lepage, P. Labelle, late D. R. Yennie,
P. Mohr and B. N. Taylor for useful discussions.
Thanks are due to M. I. Eides for communicating their preliminary result.
This research is supported in part by the U. S. National Science
Foundation.
Part of numerical work was conducted at the Cornell National
Supercomputing Facility, which receives major funding from the US
National Science Foundation and the IBM Corporation, with additional
support from New York State and members of the Corporate Research
Institute.

\appendix

\section{Projection Operators for QED Renormalization Constants}
\label{projection}

We present the projection operators for
QED renormalization constants determined by the on-shell scheme.
Our projection operators for QED renormalization
constants are quite general and  correct for any loop orders.
We set the electron mass $m=1$ through  Appendices \ref{projection} 
and \ref{KPnum}.

The projection operator of vertex renormalization terms
can be written as  \cite{kinoshitabook}
\eqn
 L=\frac{1}{4} Tr [ \Gamma_{0}(\gamma^{0}+1) ],
\endeqn
where $\Gamma_{\mu}$ is the proper vertex diagram.
Projection operators of mass renormalization term
and wave function renormalization terms can be defined similarly.
A proper self energy diagram of the 2n-th order has
the parametric representation
\eqn
\Sigma^{(2n)} = - \biggl( {-\alpha \over 4\pi }\biggr)^n (n-2)!
         {\bf F }  \int (dz)_G {1 \over U^2V^{n-1} }~.
\label{selfenergy}
\endeqn
(See Ref. \cite{kinoshitabook} for the definition of $U$, $V$,
${\bf F}$, etc.)
The projection to the mass renormalization is given by
\eqna
\delta m & = & \frac{1}{4} Tr  [ \Sigma(p)(\not p + 1)  ]|_{\not p=1}
\nonumber \\
        & = & \biggl ({\alpha \over \pi }\biggr)^n
    \int {(dz)_G \over U^2 } \sum_{m=0}^{n-1}{F_m \over U^mV^{n-1-m} },
\endeqna
where $\Sigma(p)$ is defined by (\ref{selfenergy}).
The projection operator of the wave function renormalization
constant is slightly more complicated
since it involves the derivative of the self energy
with respect to the
external momentum, namely  $p^{\mu}\partial  \Sigma(p) /\partial  p^{\mu} $.
It can be easily realized, however, in the Feynman parametric representation.
Taking the derivative of the
numerator with respect to $p$ leads to the quantity
\eqn
  {\bf E} = \sum_{\rm electron~ line ~only}  A_i {\bf F}_i~.
\endeqn
where ${\bf F}_i $ is obtained by replacing
$({\not D}_i + 1)$ by $ {\not p} $ in the electron-line operator ${\bf F}$.
The corresponding part of the wave function renormalization constant is
determined as
\eqna
 B_2({\rm numer.}) & = &
-\biggl({-\alpha \over 4\pi }\biggr)^n (n-2)!
\frac{1}{4} Tr [{\bf E(p)} \not p (\not p + 1) ]|_{\not p=1}
           \int (dz)_G {1 \over U^2V^{n-1} }
\nonumber \\
         &  =  & \biggl({\alpha \over \pi }\biggr)^n
          \int {(dz)_G \over U^2 }
         \sum_{m=0}^{n-1}{E_m \over U^mV^{n-1-m} }~.
\endeqna
The other part corresponding to the derivative  of the denominator
is similar to the mass renormalization $\delta m$ and is given by
\eqn
 B_2({\rm denom.})
           = \biggl  ({\alpha \over \pi }\biggr)^n
          \int {(dz)_G \over U^2 }
         \sum_{m=0}^{n-1}{ (n-1-m)~2~ G~ F_m \over U^mV^{n-m} },
\endeqn
where
\eqna
G & = & -{1 \over 2} p^{\nu}
      { \partial V \over \partial p^{\nu} }|_{p^2=1}
\nonumber \\
  & = & \sum_{\rm electron~ line~ only} z_i A_i ~.
\endeqna

\section{Numerical Calculation of the $\alpha(Z\alpha)$ Correction}
\label{KPnum}

Let us begin by showing the calculation of the spanning
photon diagram of Fig. \ref{KPqed}(a).
The Feynman parameters assigned to the electron-lines are
$z_1, z_2,$ and $z_3$ and assigned to the radiative photon is $z_4$.
Following the notation in Ref.\cite{CvK,kinoshitabook}, we use
the abbreviation for the sum of Feynman parameters
\eqn
    z_{12\cdots n}\equiv z_1 + z_2 + \cdots +z_n~.
\endeqn
For this diagram, $z_{1234}= 1$.
The momentum  flowing through each electron-line after integration
of photon loop momentum should be  expressed by the external momenta,
namely the electron momentum $p=(m,\vec{0})$  and  the momentum of
the external potential $q=(0,\vec{q})$.  For the electron-line $i$,
this may be written as
\eqn
   Q'_i = A_i~p~ + ~A_{iq}~q~,
\endeqn
where the $``$scalar" currents $A_i$ and $A_{iq}$ are found 
to be
\eqna
    && A_1  =  A_2=A_3=1-z_{123} U^{-1},~~~
      A_{1q}  =  A_{3q}= -z_2  U^{-1},~~~
\nonumber \\
    && A_{2q}  =  1 +A_{1q},~~~
      U=z_{1234}=1~.
\endeqna
Then the electron-line including all of numerical factors
is obtained in the operator form
\eqn
 {3 \over 8}~~ {1 \over \vec{q}\,^2 }
  [( \not D_1+1)~\gamma^{\mu}~(\not D_2+1)~\gamma^{\nu}~(\not D_3 +1) ]
     \int{(dz)_{1-4} \over U^2} {1 \over V^2}~,
\endeqn
where the $i$-th electron  line operator $D_i$ is
\eqn
D_i^{\mu}={1\over2}\int^{\infty}_{m_i^2} dm_i^2
{\partial \over \partial q_{i{\mu}} } ~.
\endeqn
Multiplying with the hyperfine splitting projection operator and 
taking the trace,
we obtain the integrand as the FORM output \cite{form}:
\eqna
      f(\vec{q}) &&  =  4~A_{1q}^2~A_{2q}\vec{q}\,^2 V^{-2}
       +  ( - 4~A_1^2~A_{2q} + 8~A_{1q} - 4~A_{2q})V^{-2}
\nonumber \\
    &&   +   B_{12} ( 8~A_{1q} + 4~A_{2q} ) (UV)^{-1},
\label{fspan}
\endeqna
where
\eqn
      B_{12}  =  B_{23} = B_{31}=1,~~~
      V  = z_{123}- z_{123}~A_1+z_4~\lambda^2
           +\vec{q}\,^2~z_2~A_{2q}~.
\endeqn
The contribution to the hyperfine splitting energy is
obtained as an integral  of the form
\eqn
\Delta E_1=  \alpha(Z\alpha) E_F
      { 1 \over \pi^2}
    \int^{\infty}_0 dq   \int{(dz)_{1-4} \over U^2} f(\vec{q}),
\endeqn
where
\eqn
   (dz)_{1-4}=dz_1dz_2dz_3dz_4\delta(1-z_{1234})~,
\endeqn
and $f(\vec{q})$ is given by (\ref{fspan}).

Next consider the vertex correction diagram of Fig. \ref{KPqed}(b).
Feynman parameters $z_1$ and $z_2$ are assigned to the electron
lines and $z_4$ to the photon line, with the constraint $z_{124}=1$.
The scalar currents are
\eqn
      A_1  =  A_2=1- z_{12}  U^{-1} ,~~~
      A_{2q}  =   -z_2  U^{-1},~~~
      A_{1q}  =  1 + A_{2q},~~~
      U=z_{124}=1
\endeqn
Then the electron-line  is
\eqn
 -{3 \over 8}~~~ {1 \over \vec{q}\,^2 }
   [ (\not D_1+1) ~\gamma^{\mu} ~(\not D_2+1)~\gamma^{\nu}
   (\not p+\not q +1)  ]
     \int^{\Lambda^2}_{\lambda^2} z_4 dm_4^2
\int{(dz)_{124} \over U^2} {1 \over V^2}~.
\endeqn
The integrand is  found to be
\eqna
      f(\vec{q})&&  =     - 4~A_{1q}~A_{2q}   \vec{q}\,^2  V^{-2}
       +    (  - 4 - 4~A_1~A_{2q} + 4~A_1~A_{1q} + 8~A_1 ) V^{-2}
\nonumber \\
      && +    B_{12} (  - 4 )(UV)^{-1},
\endeqna
where
\eqn
      B_{12}  = 1, ~~~
      V  = z_{12}- z_{12}~A_1+z_4~\lambda^2
           +\vec{q}\,^2~z_2~A_{2q}~.
\endeqn

As was described in Sec. \ref{sec:KP}, 
the charge renormalization and subtraction
of the second order anomalous magnetic moment is carried out 
by subtracting $f(\vec{q}=0)$.
The hyperfine splitting contribution from two diagrams of
the vertex type is
\eqn
\Delta E_2=  \alpha(Z\alpha) E_F
      { 2 \over \pi^2}
    \int^{\infty}_0 {dq \over (-\vec{q}\,^2)}
     \int^{\Lambda^2}_{\lambda^2} z_4 dm_4^2
         \int{(dz)_{124} \over U^2}  [f(\vec{q})-f(0) ].
\endeqn
If the $m_4^2$ integral is performed,
the denominator $V^{-2}$ becomes $V^{-1}$. The $V^{-1}$ term
is UV divergent and, if it is combined with the charge renormalization
term, it becomes $-\ln V$.
In the self-energy diagram calculation, the regularization is
implicitly performed and the resulting denominators are expressed
by $V^{-n}, n=0,1,2,\cdots $, where $V^{-0}$ implies $-\ln V $.

The last diagram is the self-energy insertion diagram
of Fig. \ref{KPqed}(c).
Feynman parameter $z_2$ is assigned to the electron-line
and $z_4$ to the photon line with the constraint $z_{24}=1$.
The scalar currents are
\eqn
      A_2  = 1- z_{2}  U^{-1},~~~
      A_{2q}  =  A_2,~~~
      U=z_{24}=1~.
\endeqn
Then the electron-line is
\eqn
 {3 \over 8}~~~ {1 \over \vec{q}\,^2 }
   [ (\not p+\not q+1) ~\gamma^{\mu}~ (\not D_2+1)~\gamma^{\nu}
      ~ (\not p+\not q+1)   ]
     \int{(dz)_{24} \over U^2} {1 \over V},
\endeqn
where
\eqn
      V  = z_{2}- z_{2}~A_2+z_4~\lambda^2
           +\vec{q}\,^2~z_2~A_{2q}~.
\endeqn
The integrand is found to be
\eqn
      f(\vec{q})   =    (  - 16 + 8~A_2 -4~A_2\vec{q}\,^2)(-\ln(V))~.
\endeqn
The mass and wave function renormalization terms can be written as
\eqn
      f_R  =  G( 16 - 8~A_2 ) \vec{q}\,^2  V_0^{-1}
      +  (  - 16 + 8~A_2 -4~A_2\vec{q}\,^2)(-\ln(V_0))~,
\endeqn
where
\eqn
      G  =  z_2~A_2,~~~
      V_0  =  z_{2}- z_{2}~A_2+z_4~\lambda^2~.
\endeqn
The hyperfine splitting contribution is given by
\eqn
\Delta E_3=  \alpha(Z\alpha) E_F
      { 1 \over \pi^2}
    \int^{\infty}_0 {dq \over (-\vec{q}\,^2)^2}
         \int{(dz)_{24} \over U^2}   [f(\vec{q})~-~f_R ]~.
\endeqn
Although this integral is analytically free from the
IR  singularity caused in the limit of the vanishing
external photon momentum $\vec{q}$, numerical integration
is very difficult in the small $q$ region, resulting
in the  poor convergence of the integral.
To avoid this numerical difficulty, we introduce another
parameter $ y$ varying from zero to one and combine
the term which is not proportional to $\vec{q}\,^2$  in the
original diagram  and the corresponding renormalization term together.
This leads to an integrand of the form
\eqn
      \tilde{f}(\vec{q})   =   G (  16 - 8~A_2 )\vec{q}\,^2  V_y^{-1}
          - 4~A_2 \vec{q}\,^2(-\ln(V)),
\endeqn
where
\eqn
      V_y  = z_{2}- z_{2}~A_2+z_4~\lambda^2
           +\vec{q}\,^2~y~z_2~A_{2q}~.
\endeqn
The corresponding renormalization term is
\eqn
      \tilde{f_R}  = G( 16 - 8~A_2 ) \vec{q}\,^2  V_0^{-1}
        - 4~A_2 \vec{q}\,^2(-\ln(V_0))~.
\endeqn
The resulting  hyperfine splitting contribution is
\eqn
\Delta E_3=  \alpha(Z\alpha) E_F
      { 1 \over \pi^2}
    \int^{\infty}_0 {dq \over (-\vec{q}\,^2)^2}
        \int^1_0 dy \int{(dz)_{24}  \over U^2}  
       [\tilde{f}(\vec{q})~-~\tilde{f_R} ]~.
\endeqn


\begin{references}

%%%% intro.tex
\bibitem{mariam}  F. G. Mariam  $et~al$., Phys.\ Rev.\ Lett. {\bf 49},
993 (1982);
E. Klempt $et~al$., Phys.\ Rev. D {\bf 25}, 652 (1982).

\bibitem{hughes} V. W. Hughes and G. zu Putlitz, Comm.\ Nucl.\ Part.\
Phys. {\bf 12}, 259 (1984).


\bibitem{KN1}  T. Kinoshita and M. Nio, Phys. Rev. Lett.
{\bf 72}, 3803 (1994).


\bibitem{fermi} E. Fermi, Z. Phys. {\bf 60}, 320 (1930).


\bibitem{CL}  W. E. Caswell and G. P. Lepage, Phys. Lett. B  {\bf 167},
437 (1986).

\bibitem{KL} T. Kinoshita and  G. P. Lepage,
in {\it Quantum Electrodynamics },
ed. by T. Kinoshita (World Scientific, Singapore,1990), pp.81 - 89.

\bibitem{positronium} P. Labelle, G. P. Lepage, and U. Magnea,
Phys. Rev. Lett. {\bf 72}, 2006 (1994).

\bibitem{SY}  J. R. Sapirstein and D. R. Yennie, in {\it Quantum
Electrodynamics}, ed. by T. Kinoshita (World Scientific,
Singapore,1990), pp. 560 - 672.

\bibitem{BE} S. J. Brodsky and G. W. Erickson,
Phys. Rev. {\bf 148}, 26 (1966).


\bibitem{thesis} M. Nio, Ph.D. Thesis, Cornell University, August (1995).

\bibitem{sapirstein2} Private communication from J. R. Sapirstein, 1995.

\bibitem{SGS} S. M. Schneider, W. Greiner and G. Soff,
Phys. Rev. A {\bf 50}, 118 (1994).

\bibitem{EKS0}  M. I. Eides, S. G. Karshenboim, and V. A. Shelyuto,
Phys. Lett. B {\bf 177}, 425 (1988); {\bf 202}, 572 (1988);
M. I. Eides, S. G. Karshenboim, and V. A. Shelyuto,
Ann. Phys. {\bf 205}, 231 (1991);  {\bf 205}, 291 (1991).


\bibitem{EKS1}  M. I. Eides, S. G. Karshenboim, and V. A. Shelyuto,
Phys.\ Lett. B {\bf 216}, 405 (1989);
M. I. Eides and V. A. Shelyuto, Phys.\ Lett. B {\bf 146}, 241 (1984).

\bibitem{KF}  A. Karimkhodzhaev and R. N. Faustov, Yad. Fiz. {\bf 53},
1012 (1991) [Eng. transl.: Sov.\ J.\ Nucl.\ Phys. {\bf 53}, 626 (1991)].

\bibitem{BF}  M. A. B. B\'{e}g and G. Feinberg, Phys.\ Rev.\ Lett.
{\bf 33}, 606 (1974);
G. T. Bodwin and D. R. Yennie, Phys.\ Rep. {\bf 43C}, 267 (1978).

\bibitem{BFer} M. A. B. B\'{e}g and G. Feinberg, Phys.\ Rev.\ Lett.
{\bf 35}, 130(E) (1974).

% new two references 
\bibitem{cage} M. E. Cage {et~al}., IEEE Trans. Instrum. Meas., IM-38,
284 (1989).

\bibitem{nez}  F. Nez $et~al$., Phys.\ Rev.\ Lett. {\bf 69}, 2326 (1992);
M. Weitz  $et~al$., Phys.\ Rev.\ Lett. {\bf 72}, 328 (1994).



\bibitem{EKS2}  M. I. Eides, S. G. Karshenboim, and V. A. Shelyuto,
Phys. Lett. B {\bf 229}, 285 (1989); {\bf 249}, 519 (1990).

\bibitem{EKS3}  M. I. Eides, S. G. Karshenboim, and V. A. Shelyuto,
Yad. Fiz. {\bf 55}, 466 (1992) [Eng. transl.: Sov. J. Nucl. Phys.
{\bf 55}, 257 (1992)];
Phys. Lett. B {\bf 268}, 433 (1991); Phys. Lett. B {\bf 316}, 631(E) (1993).

\bibitem{Eides}  M. I. Eides, S. G. Karshenboim, and V. A. Shelyuto,
Phys. Lett. B {\bf 319}, 545(E) (1993).


%%%%nrqed.tex


\bibitem{sapirstein} J. R. Sapirstein, Phys. Rev. Lett. {\bf 51}, 985 (1983).

\bibitem{persson} H. Persson, S. M. Schneider, W. Greiner,
G. Soff, and I. Lindgren, Gesellschaft f\"{u}r Schwerionenforschung
preprint, GSI-95  (1995),  to be published in Phys. Rev. Lett. 

\bibitem{nrqcd} G. P. Lepage, L. Magnea, C. Nakhleh,
U. Magnea,  and   K. Hornbostel,
Phys. \ Rev. D {\bf 46}, 4052 (1992).

\bibitem{schwinger} J. Schwinger, J. \ Math. \ Phys. {\bf 5}, 1606 (1964).


\bibitem{Fey}  R. P. Feynman, Phys. \ Rev. {\bf 76}, 769 (1949).

\bibitem{zwanziger} D. A. Zwanziger, Phys. \ Rev.  {\bf 121}, 1128 (1961).

\bibitem{pat} Private communication from P. Labelle.


\bibitem{french}  J. French and V. Weisskopf, Phys. Rev. {\bf 75},
1240 (1949).

%\bibitem{lecture} G. P. Lepage, Lecture on Lattice QCD,  
%Cornell University, Spring 1995.


%%%% kphfs.tex

\bibitem{KPoriginal} N. Kroll and F Pollock, Phys. Rev. {\bf 84}, 594
(1951), Phys. Rev. {\bf 86}, 876 (1952);
R. Karplus, A. Klein, and J. Schwinger, Phys. Rev. {\bf 84} 597 (1951).

\bibitem{yennie}  J. R. Sapirstein, E. A. Terray, and D. R. Yennie,
Phys. Rev. Lett. {\bf 51}, 982 (1983); Phys. Rev. D{\bf 29}, 2290 (1984).


\bibitem{CL2}  W. E. Caswell and G. P. Lepage, Phys. Rev. Lett. {\bf 41}
1092 (1978).

\bibitem{CvK} P. Cvitanovic and T. Kinoshita,
Phys. Rev. D {\bf 10}, 3978 (1974).

\bibitem{kinoshitabook} T. Kinoshita, in {\it Quantum
Electrodynamics}, ed. by T. Kinoshita (World Scientific,
Singapore,1990), pp. 218 - 321.

\bibitem{form} J. Vermaseren, Version 1.1, (1992).

\bibitem{lepage}  G. P. Lepage, J. Comput. Phys. {\rm 27}, 192 (1978).

%%%% knhfs.tex

\bibitem{schoon}  M. J. G. Veltman, 68000 Version, (1989).



\bibitem{numrecipes} W. H. Press, B. P. Flannery, S. A. Teukolsky,
and W. T. Vetterling, {\it Numerical Recipes}
(Cambridge University Press, 1989)

\bibitem{EKS4}  M. I. Eides, S. G. Karshenboim, and V. A. Shelyuto,
Phys. Lett. B {\bf 312}, 358 (1993); Yad. Fiz. {\bf 57}, 1240 (1994);
M. I. Eides, S. G. Karshenboim, and V. A. Shelyuto,
Yad. Fiz. {\bf 57}, 2158 (1994).

\bibitem{ES1}  M. I. Eides and V. A. Shelyuto,
Pis'ma v ZhETP {\bf 61}, 465 (1995) [ JETP Lett. {\bf 61} 478 (1995)] ;
Phys. Rev. A {\bf 52}, 954 (1995).


\bibitem{Karsh}  
S. G. Karshenboim, Zh. Eksp.Teor. Fiz. {\bf 103}, 1105
(1993) [Eng. transl.: JETP {\bf 76}, 541 (1993)];
S. G. Karshenboim,  preprint (1995), to be published 
in Zeitschrift f\"{u}r Physik D.


\end{references}
\end{document}